\documentclass[letterpaper,11pt]{article}

\usepackage[letterpaper, left=1in, top=1in, right=1in, bottom=1in,nohead,includefoot]{geometry}
\usepackage{color,charter,graphicx,natbib,enumitem,amsmath,amssymb,amsfonts,titlesec,placeins,bm,pdfpages, subfig, placeins,setspace,array}
\usepackage[utf8]{inputenc} 
\usepackage[title]{appendix}
\usepackage{sidecap} \sidecaptionvpos{figure}{c} 
\usepackage[svgnames,x11names]{xcolor}
\usepackage[pdftex]{hyperref}
	\hypersetup{
	colorlinks,%
	linkcolor=MidnightBlue,  
	urlcolor=DarkSlateGray,   
	citecolor=RoyalBlue2  
	}
\titleformat*{\section}{\normalfont\Large\bfseries\blu}
\titleformat*{\subsection}{\normalfont\large\bfseries\blu}
\titleformat*{\subsubsection}{\normalfont\normalsize\bfseries\blu}

\def\bem#1{{\blu\em #1}} 
 
\def\blu{\color{RoyalBlue4}}      
\def\bone{\mathbf{1}}\def\bzero{\mathbf{0}}

\def\H{\mathbf{H}}

\def\a{\mathbf{a}}\def\e{\textrm{e}}\def\h{\mathbf{h}}
\def\m{\mathbf{m}}\def\p{\mathbf{p}}\def\s{\mathbf{s}}
\def\x{\mathbf{x}}\def\y{\mathbf{y}}  

\def\balpha{{\bm\alpha}}
\def\btau{{\bm\tau}}

\def\cH{\mathcal{H}}\def\cS{\mathcal{S}}  

\def\seq#1#2{#1{:}#2}\def\j1J{j=\seq 1J}
\def\eqn#1{eqn.~(\ref{eq:#1})}

\def\bmat{\begin{pmatrix}}\def\emat{\end{pmatrix}}  

\def\pdfs{p.d.f.s }
\def\ci{\perp\!\!\!\perp}

\def\tE{\textrm{E}}   \def\tN{\textrm{N}}   \def\tP{\textrm{P}}\def\tPr{\textrm{Pr}}\def\tKL{\textrm{KL}}

\def\bi{\begin{itemize}[itemsep=1pt,topsep=3pt]}
\def\ei{\end{itemize}} 
\def\bna{\begin{enumerate}[itemsep=1pt,topsep=3pt,label=(\alph*)]}\def\en{\end{enumerate}}
\def\bn{\begin{enumerate}[itemsep=1pt,topsep=3pt]}\def\en{\end{enumerate}}
\def\beq#1{\begin{equation}\label{eq:#1}}\def\eeq{\end{equation}}
\def\beas{\begin{align*}}\def\eeas{\end{align*}}
\def\bea{\begin{align}}\def\eea{\end{align}}
 
\newcommand{\blind}0 

\def\scn{\cS}\def\scnj{\scn_j}
\def\KL{K\"ullback-Leibler }\def\klpf{\tKL(p\|f)}\def\klfp{\tKL(f\|p)}\def\emr{\pi_{pf}}\def\emrpf{\emr}\def\emrfp{\pi_{fp}}
\def\py{p(\y)}\def\pzeroy{p_0(\y)}\def\pjy{p_j(\y)}\def\fya{f(\y|\balpha)}
\def\alh{\widehat{\balpha}}\def\als{\balpha^*}\def\emrh{\widehat{\pi}_{pf}}\def\fy{f(\y)}\def\ky{k(\y)}\def\logpostalpha{\lambda(\balpha)} 
\def\kpf{\kappa_{pf}} \def\kfp{\kappa_{fp}} 
\def\tDir{\textrm{Dir}}
\def\safehaven{backstop }\def\safehavennospace{backstop} \def\Safehaven{Backstop }
\usepackage{tabu}
\newcolumntype{L}[1]{>{\raggedright\let\newline\\\arraybackslash\hspace{0pt}}m{#1}}
\newcolumntype{C}[1]{>{\centering\let\newline\\\arraybackslash\hspace{0pt}}m{#1}}
\newcolumntype{R}[1]{>{\raggedleft\let\newline\\\arraybackslash\hspace{0pt}}m{#1}}
 
\usepackage{arydshln} 
\usepackage{wrapfig}  
\usepackage{multirow}

\begin{document}\emergencystretch 3em  

\begin{center} 

{\blu\bf\LARGE Scenario Synthesis and Macroeconomic Risk} 

\if0\blind
	{ \bigskip
		{\large  
			    Tobias Adrian,\footnote{{\blu Tobias Adrian}, Director of the Monetary and Capital Markets Department, International Monetary Fund 
                							    \\ \indent\indent\indent  700 19th Street NW, Washington, DC 20431, U.S.A.  
		  							     \\ \indent\indent\indent \href{mailto:tadrian@imf.org}{tadrian@imf.org}}
                Domenico Giannone,\footnote{{\blu Domenico Giannone}, Assistant Director, International Monetary Fund 
                							    \\ \indent\indent\indent  700 19th Street NW, Washington, DC 20431, U.S.A.  
		  							     \\ \indent\indent\indent \href{mailto:dgiannon2@gmail.com}{dgiannon2@gmail.com}}
                Matteo Luciani,\footnote{{\blu Matteo Luciani}, Principal Economist, Board of Governors of the Federal Reserve System
	         							   \\ \indent\indent\indent 20th Street and Constitution Avenue NW, Washington, DC 20551, U.S.A.
	        							    \\ \indent\indent\indent \href{mailto:matteo.luciani@frb.gov}{matteo.luciani@frb.gov}} 
			    Mike West\footnote{{\blu Mike West}, The Arts \& Sciences Distinguished Professor Emeritus of Statistics \& Decision Sciences   
	   	  							 \\ \indent\indent\indent Duke University, Durham, NC 27708, U.S.A. 
		 							  \\ \indent\indent\indent \href{mailto:mike.west@duke.edu}{mike.west@duke.edu}}}
	} \fi  

\bigskip\bigskip
\today 
\bigskip\bigskip

\thispagestyle{empty}\setcounter{page}0

{\blu\bf \Large Abstract} 
\end{center}

\noindent We introduce methodology to bridge scenario analysis and model-based risk forecasting, leveraging their respective strengths in policy settings. Our Bayesian framework addresses the fundamental challenge of reconciling judgmental narrative approaches with statistical forecasting. Analysis evaluates explicit measures of concordance of scenarios with a reference forecasting model,   delivers Bayesian predictive synthesis of the scenarios to best match that reference, and addresses scenario set incompleteness. This underlies  systematic evaluation and integration of risks from different scenarios, and quantifies relative support for scenarios modulo the defined reference forecasts. The framework offers advances in forecasting in policy institutions that supports clear and rigorous communication of evolving risks. We also discuss broader questions of integrating judgmental information with statistical model-based forecasts in the face of unexpected circumstances.
\bigskip

\noindent {\bf Keywords:} 
\noindent Macroeconomic Forecasting, Mixtures of Scenarios, Misclassification Rates, Entropic Tilting, Bayesian Predictive Synthesis, Judgmental Forecasting, Forecast Risk Assessment

\setstretch{1.0} 
\thispagestyle{empty}
\if0\blind
	{   \renewcommand{\thefootnote}{ } 
            \footnotetext{\\[3pt] \noindent {\bf Disclaimer}: The views expressed in this paper are those of the authors and do not necessarily reflect the views and policies of the Board of Governors, the Federal Reserve System, or the International Monetary Fund, its Management, or its Executive Directors.}
            \renewcommand{\thefootnote}{\arabic{footnote}}
            \setcounter{footnote}{0}
	} \fi  
 
\newpage					        

%
\section{Introduction\label{sec:Introduction}} 
\noindent Macroeconomic policy institutions such as central banks rely heavily on forecasting methods. Monetary policymakers are regularly briefed on the economic outlook, alternative policy paths, and the balance of risks around the central forecast. Central bank staff rely on a combination of structural macroeconomic models, reduced-form empirical models, and judgmental approaches to prepare such monetary policy briefings. The central forecast is used as a basis for alternative policy path discussions, and the balance of risks is discussed more loosely based on scenario analysis. 

The Bank of England pioneered communication of risk with fan charts in 1993; the  Inflation Reports show  central projections of inflation with charts that reflect uncertainty. Uncertainty intervals are derived from judgmental assessments of risk around the baseline~\citep{BankofEngland1998}. Since 1995, the U.S. Federal Reserve's Tealbook (TB) has presented scenarios as perturbations around baseline forecasts. Most major central banks now use some variant of these approaches. 
Fan charts and scenario analysis pose practical challenges as they require frequent updating and quantification of risks based on judgment. Hence central banks are relying more often on statistical methods to forecast macroeconomic risk. The  density forecasting approach of  \lq\lq Growth-at-Risk''  (GaR:~\citealp{AdrianEtal2016,Adrianetal2019,PlagborgEtal2020,Adrianetal2022})  is increasingly popular.  The Tealbook has included GaR measures together with scenarios since 2017; other central banks have also implemented GaR approaches in addition to the more judgmental scenario methods~\citep[e.g.,][]{FIGUERES2020109126, lenza2023density, GaR_BoE2019,GaR_BoE2024, Anesti2023, BdFGaR2022,  BoIGaR2019}. 

Our focus here is on a formal statistical approach to integrating scenario-based balance of risk discussions with statistical forecasts.  The methodology defines a synthesis of the baseline and scenarios that best match the statistical {\em reference} forecast distribution, the latter typically from GaR and/or a quantile regression model.
The  scenario synthesis assigns weights to each scenario, quantifying their relative concordance  with the reference and so providing explication of why a certain set of scenarios is particularly relevant.
The analysis also incorporates a synthetic \lq\lq \safehavennospace'' scenario  designed to address potential incompleteness of the defined scenario set.
In practice,  uncertainty measures are usually published only for the baseline; alternative scenarios are typically represented only in terms of point forecasts. We use extensions of the Bayesian decision analysis method of entropic tilting~\citep[e.g.][]{RobertsonET2005,TallmanWestET2022} to define full scenario forecast distributions as perturbations  the baseline.  

Analysis further addresses scenario information beyond a single point forecast, specifically to use scenario tail percentiles that reflect measures of scenario risk. This links to the  desirability of scenario hypotheses that represent more radical perturbations of the baseline than has been 
typical~(e.g.,~\citealp{justiniano2008time, fernandez2011risk, AdrianBoyarchenko12, HeKrishnamurthy12a, BrunnermeierSannikov14, fernandez2023financial, fernandez2024search}
with structural models, and 
\citealp{adrian2021multimodality, Caldara2021, carriero2024capturing} 
with reduced-form models).
That scenarios considered by policy institutions often represent only modest perturbations of the baseline is also partly addressed by our use of the 
synthetic  \safehaven scenario. This can serve as a \lq\lq red flag"
when the scenario set fails to account for risks-- especially tail risks--  supported under the statistical reference. 

Relating baseline and scenario forecasts to the statistical reference 
exploits Bayesian predictive synthesis~\citep[e.g.][]{McAlinnWest2019,JohnsonWest2024} to motivate a discrete mixture 
(linear pool) of baseline and scenario distributions as a proxy for scenario-based forecasting. The \lq\lq match" of such a mixture with the statistical reference distribution uses a central measure of concordance between distributions, namely the {\em expected misclassification rate} (EMR).  Identifying mixture weights to optimize EMR is then a formal Bayesian decision problem. Relative scenario probabilities based on this scenario:reference optimization guide evaluation and interpretation of the roles of scenarios. The analysis includes explicit statistical measures of {\em scenario set incompleteness} reflecting aspects of lack-of-concordance the scenario synthesis with the reference.  This aids in the policy setting on the question of whether the baseline and chosen scenarios adequately reflect all the risks captured by the reference.  

The case study draws on published versions of the TB. We use data from reports prepared for the December FOMC meetings in 2007 and 2018, giving predictions for 2008 and 2019, respectively. Following the TB, we  focus on risks to real growth. Reanalysis incorporating other variables, such as inflation and unemployment at risk~\citep{AdamsEtal2021,Kiley2022,Loria2024Inflation} is straightforward but beyond our main scope here. 
Our detailed examples highlight the generation of scenario weights reflecting aspects of concordance with the reference, 
and also the questions of scenario set incompleteness. One example of that latter highlights the lack of a very negative,  \lq\lq downside risk scenario"  in both the 2007 and 2018 TB.  This relates to the particular interest in our analysis 
when economic uncertainty is high so that defining an adequate  baseline forecast is challenging. Then, listing and discussing a range of plausible scenarios, each with an assigned probability derived from the reference match, offers 
a richer perspective on informed decision-making under uncertainty.

Our analysis takes baseline and scenarios (as well as the reference) as given. In policy practice, of course, the back-and-forth between changes to statistical forecast distributions and the evolving narrative of scenarios provides a rich ground to rigorously examine shifts in the balance of risks. This was noted by \cite{Bernanke2023} and is germane to the TB, where scenario-based approaches to the balance of risk and statistical forecast distributions are discussed separately. As Federal Reserve Chair Jerome Powell noted during the Press Conference following the January 2025 FOMC meeting, {\em \blu \lq\lq One of the things our staff does is they look at a range of possible outcomes. [...] There’ll be baseline, and then they’ll show six or seven alternative scenarios, including really good ones and not so good ones. And what those do is they spark [...] the policymakers to sort of think and understand about [...] the uncertainties that surround us.''} 
Our methodology provides formal cross-talk that can aid macroeconomic staff in policy institutions: it combines the communicative strength of narrative scenarios with the statistical rigor of predictive models, 
identifying the most relevant risks with easy-to-understand stories and quantifying the relevance of these stories.


Section~\ref{sec:context} discusses foundations and overviews   methodology. Section~\ref{sec:scenarioET} addresses  partial scenario information. Section~\ref{sec:EMRetc} introduces expected misclassification rates as distributional concordance metrics, with foundational insights. Section~\ref{sec:Methodology} develops the embedding of scenario analysis in a fully Bayesian framework, with core theoretical summaries and aspects of computational implementation.  Section~\ref{sec:empirics} summarizes key aspects of the detailed 
case study. 
Section~\ref{sec:judgmentaltilting} links to broader questions of combining  judgmental information with statistical model-based forecasts. The Appendix adds technical and methodological details. Summary comments define Section~\ref{sec:conclusions}.

%
\newpage
\section{Setting, Foundations and Perspective \label{sec:context}} 
\subsection{Context and Goals\label{sec:settinggoals}}

Interest lies in forecasting a vector outcome $\y$, such as a path of several macroeconomic indicators over multiple future time periods, based on the following ingredients.
\bi
\item A policy-based analysis produces a predictive density $\pzeroy$, referred to as the \bem{baseline density}.   
\item Relative to the baseline, the policy analysis considers each of a set of \bem{alternative scenarios}; scenario $j$, labeled $\scnj$, generates a predictive density $p_j(\y)$.   These are regarded as hypothetical scenarios to be assessed relative to the baseline. 
\item The baseline is a given forecast distribution in the policy setting, so not an hypothetical scenario; that understood, we use $\scn_0$ and the index 
$j=0$ to designate the baseline. 
\item Separately, a statistical model  (e.g., the statistical GaR analysis) produces a full predictive density $\py$, referred to as the \bem{reference predictive density}.   
 \ei
The over-arching goal is to identify \lq\lq closeness" of each scenario to the reference $\py,$ and rank them relative to that assessment. 
The methodology we introduce addresses this, building on foundational statistical concepts and model developments now discussed.

\subsection{Scenario Mixtures and Bayesian Predictive Synthesis\label{sec:scenariomixturesBPS}}

A Bayesian decision-maker in the policy setting can regard the set of scenario p.d.f.s $p_j(\y)$ as \lq\lq information" to use in forming a policy-relevant overall forecast.  This involves some form of \bem{pooling} of the predictions across the baseline and alternative scenarios.   Here the 
foundational theory of Bayesian predictive synthesis (BPS)-- and the specific class of \lq\lq mixture BPS'' models~(\citealp{McAlinnWest2019}, section 2.2;~\citealp{JohnsonWest2024})-- applies.   Under BPS, a valid Bayesian predictive analysis can be based on a \bem{scenario mixture}, i.e., a distribution with p.d.f. 
$f(\y|\balpha)$  that is a linear pool of the $p_j(\y)$ with respect to probability weights $\alpha_j$ in a vector $\balpha,$  namely
$f(\y|\balpha) \propto \sum_j \alpha_j p_j(\y).$ 

A key theoretical aspect of mixture BPS is that it can address the broad question of \lq\lq scenario set (in-)completeness".   That is, a setting in which the baseline $\scn_0$ and all of the the alternative scenarios $\scnj$ considered are discordant with the reference $p(\y).$  This relates to the \lq\lq model set incompleteness'' issue widely discussed in Bayesian econometrics. BPS theory addresses this by requiring an additional p.d.f. to extend the initial set and to use in the mixture. This has been exploited in BPS applications-- and in its generalization to decision-guided settings (BPDS:~\citealp{TallmanWest2023,ChernisTallmanKoopWest2024})-- by structuring  the additional p.d.f. as a \bem{\safehavennospace} that can be expected to be supported by future data that is not so well-predicted by the initial model set.
Examples in the above studies use an over-dispersed average of the initial mixture of model p.d.f., and this strategy can be adopted for scenario analysis. 
The specific construction of such a \safehaven scenario in the case study in Section~\ref{sec:empirics} provides an example of this modelling strategy.

Index the alternative scenarios from the policy setting by $j=1:J-1$ with the baseline $j=0$ and now with $j=J$ for the chosen \safehaven p.d.f. The latter is labeled $\scn_J$ though it is a purely synthetic scenario chosen for the above purposes.  Then the overall \bem{scenario mixture} p.d.f. is 
\begin{equation}\label{eq:scenariomixturef} 
f(\y|\balpha) = \sum_{j=\seq 0J} \alpha_j p_j(\y).
\end{equation}

\subsection{Incomplete Specification of Scenario Forecast Distributions\label{sec:scenincompleteness}}

Scenario p.d.f.s $p_j(\y)$ are typically only partially specified.  A  common setting is that $\scnj$ defines point forecasts such as means or medians, with or without uncertainty measures such as a few other percentiles.  The foundational concept is that the alternative scenarios represent economically relevant \lq\lq what-if?'' perturbations of the baseline.  Hence receiving  such partial information on $\scnj$  indicates a modification of $p_0(\y)$ to match that partial information. Our approach aims to identify $p_j(\y)$ that is \lq\lq closest to" the baseline $p_0(\y)$ subject to being consistent with that partial scenario information. The theoretical basis for methodology to do this, detailed in 
 Section~\ref{sec:ET},  is that of entropic tilting (ET:~\citealp{TallmanWestET2022}). 
Since its introduction by~\cite{RobertsonET2005},  ET--based methodology has seen increasing use in forecasting in econometrics, finance and related areas~\citep[e.g.][]{KrugerET2017,MetaxoglouPettenuzzoSmith2018,KoopMcIntyreMitchell2019,AntolinDiazPetrellaRubioRamirez2021,ClarkGanicsMertens2022,West2023constrainedforecasting,crump2021large}.  
The current setting is different, though use here of ET is close in spirit and goals to 
its original use in imposing constraints on a given-- here the baseline-- forecast distribution.

\subsection{Scenario-Reference Concordance\label{sec:scenrefconcordancegoal}} 

The goal of measuring concordance of scenarios with the statistical reference is now that of relating $f(\y|\balpha)$ in~\eqn{scenariomixturef}
to the reference p.d.f. $\py$. This is addressed by identifying 
the probability vector $\balpha=(\alpha_0,\ldots,\alpha_J)'$ such that 
the scenario mixture is \lq\lq closest to" $p(\y).$ 
This requires specification of a utility function to characterize and quantify \lq\lq close" in comparing densities, and then the resulting methodology to evaluate $\balpha$ and thus define both scenario-specific weights and the overall mixture synthesis.  Section~\ref{sec:concordanceEMR} introduces a foundational metric for this-- based on a measure of concordance of $f(\y|\balpha)$ and $p(\y)$ from 
traditional statistical classification. With some new and relevant theoretical results and motivating examples, this underlies its use in scenario synthesis.

%
\section{Partial Scenario Information and Entropic Tilting\label{sec:scenarioET}}
  
\subsection{Partial Scenario Information\label{sec:scenarioinformation}}

As noted in Section~\ref{sec:scenincompleteness} the common setting is that for each scenario only partial information relative to the fully specified baseline is provided. In many  examples, the partial information can be represented as expectations of functions of $\y$, and this is the setting we adopt. Often, only the perturbed central tendency is reported. If taken as a mean, it would be a constraint on the expected value directly. If taken as the median, then it is formally defined as the expectation of an indicator function. Similar reasoning applies to other percentiles. Our analysis below addresses multiple  scenario features simultaneously,  such as a set of percentiles.  

Suppose that $\scnj$ provides partial information on $p_j(\y)$ in terms of $\m_j = \tE[\s_j(\y)|\scnj]$ where $\s_j(\y)$ is a $q_j-$vector of \bem{scenario scores}; call the given vector $\m_j$ the \bem{target score} for $\scnj$. In general, the definition of scores can be scenario-specific, but here we assume that $q_j=q$ and $\s_j(\y)=\s(\y)$ for all $j=\seq 1J$. A main case of interest has elements of $\s(\y)$ as indicator functions in one or more of the univariate dimensions; then  $\m_j$ is a given vector of percentiles of $p_j(\y)$ in those dimensions.   With $\scnj$ regarded as a perturbations of the baseline, methodology aims at identifying that $p_j(\y)$ closest to the baseline $p_0(\y)$ subject to being consistent with the forecast information $\m_j$.  Entropic tilting (ET) results if we choose to define \lq\lq close to" in a K\"ullback-Leibler (KL) sense.  

\subsection{Entropic Tilting and Scenario-Baseline ET Weights\label{sec:ET}}

ET--based methodology, recently exploited in new ways in Bayesian predictive decision synthesis~\citep[e.g.][]{ChernisTallmanKoopWest2024, TallmanWest2023,TallmanWest2024}, was originally used in imposing constraints on forecast distributions; that is the context here. 
In our setting, ET aims to identify $p_j(\y)$ to minimize the KL divergence of the baseline 
$p_0(\y)$ from $p_j(\y)$ subject to   $\m_j = \tE[\s_j(\y)|\scnj] = \int_\y \s_j(\y)p_j(\y)d\y$.
ET theory~\citep{TallmanWestET2022} yields
\beq{ETpj} p_j(\y) = k_j \e^{\btau_j'\s_j(\y)} p_0(\y)\quad\textrm{where}\quad k_j^{-1} =  \int_\y \e^{\btau_j'\s_j(\y)} p_0(\y)d\y,\eeq
in which $\btau_j$ is the (provably unique) \bem{tilting vector} such that the expectation constraint is satisfied.  
The implied identity  $\bzero  =  \int_\y \{\s_j(\y)-\m_j\} \textrm{exp}\{\btau_j'\s_j(\y)\} p_0(\y)d\y$ 
is typically efficiently solved for  $\btau_j$  using simple Newton-Raphson. 

In practice, it is typical that the baseline is represented in terms of a Monte Carlo (MC) sample, i.e., defined as a discrete distribution $\{ \y^i, w_0^i \}_{i=\seq 1n}$ with support points $\y^i$ having weight (probability) $w_0^i.$ This is particularly key in our setting as we will later use importance sampling to evaluate $p_0(\y)$ relative to the statistical reference $p(\y)$.  Then expectations defining the ET tilting vectors $\btau_j$ are trivially evaluated via simple Monte Carlo integration. 

ET analysis can be regarded as using $p_0(\y)$ as an importance sampling proposal with respect to a target p.d.f.  $p_j(\y).$ This was recognized by~\cite{RobertsonET2005} and provides useful numerical checks on consistency of the scenario-specific moment constraints with the baseline.   On sample values $\y^i$, the implied normalized IS weights for MC integration in \eqn{ETpj} are  $w_j^i \propto u_j^i w_0^i$   
where  $u_j^i\propto p_j(\y^i)/p_0(\y^i) = \textrm{exp}\{\btau_j'\s_j(\y^i)\}$.   The $u_j^i$ are called \bem{ET weights}. The standard expected sample size (ESS) can be evaluated on the $u_i$. ESS-- the reciprocal of the sum of squared $u_j^i$ over $i=\seq 1n$-- provides an overall assessment of concordance of the $\scnj$ constraints with the baseline~\citep[][sect.~1.6]{TallmanWestET2022}. This relates closely to the minimized KL  divergence~(e.g.,~\citealp[][sect.~3.3]{GruberWest2016BA}; \citealp[][sect.~5.4]{GruberWest2017ECOSTA}) but on an interpretable scale.

\section{Predictive Concordance\label{sec:EMRetc}}

\subsection{Predictive Concordance and Misclassification Rates\label{sec:concordanceEMR}}

Predictive concordance mooted in Section~\ref{sec:scenrefconcordancegoal}
is presented here in a general setting comparing two density functions $\py$ and $\fy.$   The scenario mixture setting then arises with $\fy$ replaced by $\fya$ of \eqn{scenariomixturef} for any given $\balpha.$ Assume that $\py$ and $\fy$ have the same support.

Suppose a random draw $\y$ is made from either $\fy$ or $\py$ with equal probabilities. It is not disclosed which distribution generates the outcome $\y.$ Write $\cH_p$ for the hypothesis that $\y\sim\py,$ and $\cH_f$ for the hypothesis that $\y\sim\fy.$ 
Since the choice is made with $\tPr(\cH_p)=\tPr(\cH_f)=0.5,$ the resulting
posterior probabilities conditional on the observed $\y$ are   
$ \tP(\cH_p|\y) = \py/\{\py+\fy\}$ and $\tPr(\cH_f|\y)=1-\tP(\cH_p|\y).$ 

Now assume that $\y$ is actually a draw from $\py$, i.e., condition on $\cH_p$.   
Before learning $\y,$ the expected posterior probability on $\cH_f$ is then 
\beq{Eppyf} 
\emr \equiv \tE[\tP(\cH_f|\y) |\cH_p] = \int_\y  \tP(\cH_f|\y) \py d\y =  \int_\y  \frac{\fy\py}{\{\fy+\py\}}d\y.
\eeq
By symmetry,  if $\y$ is actually from $\cH_f,$ the 
expected posterior probability $\emrfp= \tE[\tP(\cH_p|\y) |\cH_f]$ is obviously the same, $\emrfp = \emr$.
 
Predictive concordance of $\fy$ with $\py$  is inherently measured by the \bem{expected misclassification rate (EMR)} $\emr$. Higher values indicate that  it is difficult to discriminate  $\fy$ from $\py$-- indicating that draws from $\fy$ are more likely to be misclassified as coming from $\py$--  and vice-versa.   This is a natural, interpretable  metric to assess concordance-- or discordance-- of the two distributions.  

In traditional classification in statistics and machine learning, the optimal Bayesian classifier judges $\y$ as coming from $f(\y)$ with probability $\tP(\cH_f|\y).$ Averaging across $\y\sim p(\cdot),$ and using standard terminology,  
$1-\pi_{pf}$ is then both the population {\em sensitivity} 
and (due to the comparison of just two distributions and the implied symmetry) 
the population {\em sensitivity}  
of the optimal Bayesian classifier. It follows that $1-\pi_{pf}$ is the  traditional overall {\em accuracy} of the test comparing $f(\cdot)$ and $p(\cdot)$, and so EMR $\pi_{pf} =1-\textrm{\em accuracy}$ is the traditional {\em error rate}. Increasing EMR indicates decreased discrimination of $f(\cdot)$ from $p(\cdot)$. Judging $f(\cdot)$ to be
\lq\lq close to" $p(\cdot)$ at higher values of $\pi_{pf}$ is thus theoretically fundamental and practically interpretable.  
 
It is immediate that $\emr\le 0.5$ with equality only when $f(\cdot)\equiv p(\cdot),$ defining the absolute scale for assessment of concordance.  To prove this, note that  
$\emr = \tE[r(\y)/\{1+r(\y)\} |\cH_p]$ where $r(\y)=\fy/\py$ with $\tE[r(\y) |\cH_p]=1.$ 
Now, $r/(1+r)$ is concave on $r>0$ so   that  
$\emr \le \tE[r(\y)|\cH_p]/\{1+\tE[r(\y)|\cH_p]\}=1/2.$  
The upper bound is achieved when $f(\y)\equiv p(\y)$, i.e., $r(\y)=1$ for all $\y.$

 Now consider a decision setting where $f(\cdot)$ is to be chosen to be \lq\lq close to" $p(\y),$ and
when $\y\sim p(\cdot)$.  Choosing  $f(\cdot)$ to
maximize $\pi_{pf}$ subject to relevant constraints is the optimal decision with respect to the implied constrained version of utility function $\tP(\cH_f|\y)$.  This defines the Bayesian foundation of use of EMR  in the scenario synthesis development in Section~\ref{sec:Methodology}. 

\subsection{Relationships to K\"ullback-Leibler Divergence  \label{sec:EMRandKL}}

Note that 
$\emr = \tE[ 1/[1+\exp\{\ky\}]|\cH_p]$ where $\ky=\log\{\py/\fy\}$.
Under $\cH_p,$  the scalar random quantity $k(\y)$ has expectation $\klpf \equiv  \tE[k(\y) |\cH_p]  =  \int_y \log\{\py/\fy\}\py d\y$,  the \KL divergence {\em of} $f(\cdot)$ {\em from} $p(\cdot).$  Assuming this expectation is finite, 
the delta approximation yields $\emr\approx 1/[1+\exp\{\klpf\}]$; thus choosing $f(\cdot)$ to maximize $\emr$ is 
 approximately the KL divergence minimizing solution.  
In many cases of practical relevance, this also provides a strict lower bound 
on $\emr,$ i.e.,  $\emr \ge 1/[1+\exp\{\klpf\}];$ see Appendix~\ref{app:KLboundEMR}.   
Both the direct approximation and the lower bound  are accurate in cases of higher concordance.  
Then, the symmetry of EMR in $f(\cdot)$ and $p(\cdot)$ implies that the same results hold with the two densities exchanged.  
With $\klfp$ the divergence of  {\em of} $p(\cdot)$ {\em from} $f(\cdot),$    this immediately refines the lower bound 
to $\pi_{pf} \ge 1/[1+\exp(\kpf)]$ where $\kpf= \min\{ \klpf,  \klfp \},$  with equality as the direct delta approximation.    
KL divergence always raise the question of directional definition. 
This does not arise in using $\pi_{pf}$ due to its symmetry, 
and this link to KL indicates the relevant \lq\lq symmetrization'' of KL as $\kappa_{pf}.$
In cases of relatively good concordance, the two directional measures will also be close.  Additional aspects of the relationship are discussed and exemplified in Appendix~\ref{app:moreEMRandKL}.

EMR is fundamental for reasons discussed above; we have presented these connections 
to KL as it is a well-known measure.  A major caveat is that it assumes KL measures are finite. There are important practical contexts where this is not so. An example has $\fy$ Gaussian and $\py$ log T with any degrees of freedom; then $\py$ has no moments at all~\cite[e.g.][Supplementary Material, Appendix B]{West2023constrainedforecasting} and $\klpf$ is infinite. In contrast, $\emrpf\in (0,0.5]$ always.


\section{Scenario Synthesis\label{sec:Methodology}}

\subsection{EMR and Optimizing Scenario Mixture Probabilities\label{sec:scenmixprobabilities}} 
 
The predictive concordance concept applies to the scenario mixture setting with $\fy$ replaced by $\fya = \sum_{j=\seq 0J}\alpha_j\pjy$ at any chosen probability vector $\balpha=(\alpha_0,\ldots,\alpha_J)'.$  Making dependence on $\balpha$ explicit,~\eqn{Eppyf} is now
\beq{Eppyfalpha} 
\emr(\balpha) =  \int_\y  \frac{\fya\py}{\{\fya+\py\}}d\y.
\eeq
Values of  $\balpha$ yielding high values of 
$\emr(\balpha)$ define mixtures of baseline and scenarios \lq\lq close to'' to the reference $\py$  in terms of probabilistic concordance. 
Suppose $\alh$ maximizes $\emr(\balpha)$  with maximum value $\emrh=\emr(\alh)$.
Each element $\widehat\alpha_j$ of $\alh$ quantifies the extent to which $\scnj$ is concordant with the reference {\em\blu relative to the other $\scn_i$ for $i\ne j$}. 
The summary $\emrh$  is a concrete measure of the concordance of the set of predictions from the baseline and the scenarios combined. 
A low value of $\emrh$ indicates that 
none of the $\scnj$  nor their mixture are really concordant with the reference, relating to the scenario set incompleteness discussion of 
Section~\ref{sec:scenariomixturesBPS}.
Thus $\emrh$ measures how \lq\lq discordant'' the scenario set is with the reference statistical predictions. The weight $\widehat\alpha_J$ on the \safehaven provides additional information.

The framework addresses selection of $\balpha$ as a decision problem that maximizes 
$\emr(\balpha)$ with regularization to penalize very small $\alpha_j.$ This is based on deeper foundational and theoretical development in the next subsection, and leads to choosing $\als$ to maximize the objective function
\beq{Logpostalpha} 
\logpostalpha =  \log\{\emr(\balpha)\} + \epsilon\sum_{j=\seq 0J} \log(\alpha_j), \quad\textrm{subject to } \alpha_j>0 \ (j=\seq 0J) \quad\textrm{and  } \sum_{j=\seq 0J}\alpha_j=1,
\eeq
where $\epsilon>0$ is a very small regularization parameter. As we now show,~\eqn{Logpostalpha} is in fact the log of a formal posterior distribution so that the optimization seeks the posterior mode.

\subsection{Bayesian Foundation\label{sec:priorposteriorEMR}}

\subsubsection{EMR is a Likelihood Function\label{sec:EMRlikelohood}}

Suppose that the economic reality $\y$ is generated from the reference $p(\y)$ and 
consider an hypothetical/synthetic binary outcome $z$ generated from the Bernoulli distribution 
with success probability $\tPr(z=1|\y,\balpha) = f(\y|\balpha)/\{p(\y)+f(\y|\balpha)\}.$ Then
$$p(z=1,\y|\balpha) = \tPr(z=1|\y,\balpha) p(\y|\balpha) = \tPr(z=1|\y,\balpha)p(\y) 
=  f(\y|\balpha)p(\y)/\{p(\y)+f(\y|\balpha)\}.$$ 
Now suppose you observe $z=1$ but not $\y$; EMR emerges via expectations over the \lq\lq missing data" $\y,$  viz., $p(z=1|\balpha) = \emr(\balpha).$   Thus, $\emr(\balpha)$ is in fact a likelihood function for the {\em parameter} $\balpha$ based on an hypothetical observation $z=1$ that classifies a random draw from $p(\y)$ as coming from $\fy$ under a 50:50 prior.  The connection with the foundation of EMR in Section~\ref{sec:concordanceEMR} is immediate.

It follows that EMR-maximizer $\alh$ is a maximum likelihood estimate (MLE). 
Evaluating $\alh$ is probability simplex constrained convex optimization problem with a unique solution, the convexity and hence uniqueness being shown here in Appendix~\ref{app:MLEconvex}.  
The solution will typically be a \bem{sparse mixture} of scenarios, with some zeros in $\alh.$  This follows
from general results of optimization of convex functions over the probability simplex~\citep[e.g.][]{BoydVandenberghe2004}.
For some integer $k\in \{0:J\}$ a subset of $k$ of the $\widehat\alpha_j$ can be zero. There are cases when $k=0$ but $k>0$-- defining a sparse optimizing vector-- is more usual, especially with larger $J$ and diversity among the $p_j(\y).$ 
This relates to general features of optimization over the simplex; simplex constraints operate to shrink weights to the boundaries, effectively as  $\ell_1$ shrinkage for sparsity~\cite[e.g.][]{brodie2009sparse}. This underlies the notion of scalability of the analysis to larger numbers of scenarios.

However, sparsity in $\alh$ is unstable since it is not a genuine feature but  is induced by the implicit prior $\ell_1$ penalty; its values are typically very sensitive to small changes in the input scenario and reference p.d.f.s. This pathology of sparsity inducing penalties was identified and documented in the context of forecasting by \citet{IllusionOfSparsity}. In the current setting, take an example with two very similar scenario p.d.f.s; one of these scenarios will have a zero value in $\alh,$ the other non-zero.  Then, a very small change in either of the p.d.f.s-- or of the reference p.d.f.-- will flip the zero/non-zero pattern.  At each of these extremes-- and for ranges of the $\alpha_j$ on these two scenarios 
bridging the extremes-- the resulting scenario mixture $f(\y|\alh)$ will be almost unchanged. This  sensitivity is undesirable; it is desirable to have similar probabilities on the two scenarios. The key point is that a uniform prior on $\balpha$ favors overly sparse models when the likelihood function has modes at the simplex boundaries. This can be addressed by imposing additional constraints or, more foundationally, with a minimally informative \lq\lq regularizing" prior over $\balpha$.

\subsubsection{Priors and Penalties\label{sec:prioralpha}}

The natural priors are Dirichlet, $\balpha\sim \tDir(\a)$ having p.d.f. 
$p(\balpha) \propto \prod_{j=\seq 0J} \alpha_j^{a_j-1}$ over the simplex. Here $a_j>0$ for all $j$
and, with precision $a=\sum_{j=\seq 0J}a_j,$ the means are $a_j/a$ and 
prior joint mode has elements $\max\{0, (a_j-1)/(a-J-1)\}.$ A prior with each $a_j=1+\epsilon$ for a very small $\epsilon>0$ is \lq\lq minimally informative" subject to the joint prior mode being positive on each scenario.   Modifications to $a_j=1+\epsilon_j$ to differentially favor scenarios {\em a priori} are obviously of interest,  but for this paper the symmetric prior is adopted. For given $\epsilon,$  the prior joint mode and mean are then each $\bone/(J+1),$  i.e., favoring a uniform set of scenario probabilities though with high uncertainty since $\epsilon$ is taken as very small.  Under this prior $\balpha\sim \tDir(\bone(1+\epsilon)),$  the log posterior is $\logpostalpha$ in~\eqn{Logpostalpha}, up to an  additive constant. The prior is zero at simplex boundaries, hence so is the resulting unimodal posterior.  The posterior mode-- denoted by $\als$-- maximizes EMR modified by the prior-based penalty that explicitly acts to move from the boundary zero MLE values in $\alh$ to small but non-zero values. This leads to more stable and robust results and addresses the issues discussed in the previous section.

Analysis requires choice of a (small) value of 
the regularizing hyper-parameter $\epsilon.$ Based on theory in Appendix~\ref{sec:calibrateepsilon}, the default recommendation is $\epsilon = c/(J+1),$ where $c=0.005.$ The value of $c$ can be modified somewhat up/down with minimal impact, while the scaling with number of scenarios is important in more heavily penalizing the MLE-based analysis in higher dimensions.  Given $\epsilon>0,$ evaluation of the posterior mode $\als$ to maximize~\eqn{Logpostalpha} trivially modifies the probability simplex constrained convex optimization problem with a unique solution.  See  Appendix~\ref{app:MLEconvex}.

It is also of interest to consider analyses with additional constraints on $\balpha.$ 
A key example is to require $\alpha_0\ge \alpha_j$ for $j=\seq 1J,$ consistent with the view that 
the baseline is the \lq\lq modal" scenario. In general it is of interest to run comparative analyses with and without such constraints. Such a constraint simply modifies the Dirichlet prior by the indicator of the constraint; this does not impact the convexity of the optimization problem and is trivially implemented.

\subsection{Monte Carlo Importance Sampling\label{sec:evaluatepigivenalpha}}  
  
Analysis {\em prima facie} relies on evaluating the \pdfs $\py$ and each $\pjy,$  and then performing the integration in \eqn{Eppyfalpha}.
Analytic approximations to the integral may be explored. Specific approximations relate to measures of discriminatory information in classification using mixtures~\citep[e.g.][]{LinChanWest2015Biostatistics}.  
In practice, however-- and as already noted in Section~\ref{sec:ET}-- forecasts will typically be \lq\lq available'' in terms of Monte Carlo (MC) samples, so direct evaluation of $\pi_{pf}$ by MC integration is a priority (and avoids concerns of assessing the quality of analytic approximations).

The analysis is implemented with the values of the \pdfs $\py$ and the $\pjy$ available only on a (large) sample of MC draws from the reference $\py$, a \bem{reference random sample}.  The random sample $\y^i$, $(i=\seq 1n)$, is drawn from $p(\y)$  and at the first step this defines an importance sample (IS) for the baseline $p_0(\y)$ with normalized IS weights  $w_0^i \propto p_0(\y^i)/p(\y^i).$  The discrete distribution $\{ \y^i, w_0^i \}_{i=\seq 1n}$  defines the MC approximation to the baseline for evaluation of expectations in the downstream analysis.  A proviso is that $\py$ is a relevant importance sampling proposal; in particular, it should be heavier-tailed than $p_0(\y)$.  As in all applications of IS, monitoring efficiency measures such  as the \% effective sample size 
$\textrm{ESS}=n^{-1}100/\sum_{i=\seq1n} (w_0^i)^2$ provides guidance; initial analysis generating a relatively low ESS guides choice of a larger sample size. 
This IS analysis then underlies evaluation of scenario-specific ET parameters as in Section~\ref{sec:ET}, 
yielding ET weights $u_j^i\propto \textrm{exp}\{\btau_j'\s_j(\y^i)\}$ on sample $\y^i$ defining   $p_j(\cdot)$ relative to the baseline. Scenario-specific ESS measures using the $u_j^i$ weights are then relevant.  In (rare) cases of a scenario that is really discordant with the baseline, a very low ESS indicates such.  Refined but much more computationally demanding adaptive IS methods may be considered, but are outside our current scope.  In any case, encountering such discordance would indicate that such a scenario might better 
be considered separately and its full distribution directly assessed.

This ET analysis leads to {\em compound weights} $w_j^i \propto u_j^iw_0^i$   relating $\scnj$ to the reference; these are called the \bem{ET$-$IS weights}. ESS measures can now also be evaluated on the $w_j^i$ to provide direct overall assessment of each $p_j(\cdot)$ relative to the reference $p(\cdot).$  Note that there can be cases where a scenario is more concordant with the reference than the baseline as some of our examples show.    The reference sample and compound ET$-$IS weights are then ingredients in the direct 
evaluation of~\eqn{Eppyfalpha} via MC integration.

%
%
\section{Case Study\label{sec:empirics}}
The case study draws from the Risk and Uncertainty analyses in the 
December \href{https://www.federalreserve.gov/monetarypolicy/files/FOMC20071211gbpt120071205.pdf}{2007} and \href{https://www.federalreserve.gov/monetarypolicy/files/FOMC20181219tealbooka20181207.pdf}{2018} Tealbooks \citep{Dec2007Tealbook,Dec2018Tealbook}. The scenarios specify point forecasts for GDP growth, inflation, the unemployment rate, and other variables.  The methodology applies to multiple variables and horizons, but this first application
restricts attention to one-year ahead GDP growth, namely $\y=y$, now scalar. Analysis follows the processes discussed in the previous sections; Appendix~\ref{sec:summaryofflow} gives a summary of the flow of analytic 
 and computational details. 


%
\subsection{Reference Distribution\label{sec:ReferenceInfo}}

Among recent statistical approaches to risk assessment,~\cite{Adrianetal2019} develop quantile regression models of conditional predictive distributions and show that financial markets provide useful risk information. This approach has influenced practice, being adopted for conditional one-year ahead forecasts of GDP growth, unemployment rate, and inflation, for example, by the Federal Reserve Board~\citep{TBTVMR} in the \lq\lq Time-Varying Macroeconomic Risk'' exhibit in the Risk and Uncertainty section of Tealbook A,  and the New York Federal Reserve~\citep{Outlook@Risk} in \href{https://www.newyorkfed.org/research/policy/outlook-at-risk}{Outlook-at-Risk}. Other central banks and international financial institutions have also adopted GaR approaches~\citep[examples include][]{FIGUERES2020109126, lenza2023density, IMF2017, GaR_BoE2024, Anesti2023, BdFGaR2022,  BoIGaR2019, BancoEspana2022, Bundesbank2023}.

\begin{wraptable}{r}{.5\textwidth}\vspace{-1pt}\caption{Reference percentiles} \label{tab:Reference} \centering \smallskip
\setlength{\tabcolsep}{.0\textwidth} \small
\begin{tabular}{C{.02\textwidth}L{.07\textwidth}C{.09\textwidth}C{.09\textwidth}C{.01\textwidth}|C{.02\textwidth}L{.07\textwidth}C{.09\textwidth}} \hline  \hline
&		&	\multicolumn{2}{c}{NY Fed}	&\multicolumn{2}{c}{ } &	& {Tealbook}	\\  \hline 
&		&	\blu 2007	&	\blu 2018	&&&		&	\blu 2018	\\ \hline
&	\blu P10:	&	$-$1.7\phantom{\ \ \,}	&	0.0	&&&	\blu P5	&	0.7	\\
&	\blu P25:	&	0.2	&	1.1	&&&	\blu P15	&	1.3	\\
&	\blu P50:	&	1.8	&	2.1	&&&	\blu P50	&	2.5	\\
&	\blu P75:	&	3.3	&	3.0	&&&	\blu P85	&	3.6	\\
&	\blu P90:	&	4.8	&	4.0	&&&	\blu P95	&	4.3	\\ \hline
\end{tabular}
\begin{tabular}{p{.46\textwidth}}\scriptsize
The NY Fed Outlook at Risks gives point forecasts for one-year ahead GDP growth.  The Tealbook Time-Varying Macroeconomic Risk gives one-year ahead Tealbook forecast errors in the December 2018 Tealbook, which provides GDP growth point forecasts once the baseline forecast is added.
\end{tabular}\vspace{-1pt}
\end{wraptable}

The Federal Reserve Board started producing the Tealbook Time-Varying Macroeconomic Risk forecasts in 2017 but do not provide past values. The NY Fed started producing the Outlook-at-Risk forecasts only in 2023 but provides past values starting in 1989. As a result, our case study constructs and compares reference distributions from both the NY Fed and the Fed Board. In practice, both the NY Fed and the Tealbook give five predictive percentiles (Table~\ref{tab:Reference}). To construct the reference distribution, we fit skew-t distributions~\citep{AzzaliniCapitanio2003} on these percentiles. Following~\cite{Adrianetal2019}, the four parameters of the skew-t minimize the squared distance between the given reference quantiles and those of the resulting skew-t. Table~\ref{tab:refandbaselineparameters} reports resulting parameters.
\subsection{Baseline Distribution\label{sec:BaselineScenariosInfo}}
Table~\ref{tab:TBbaseline} shows baseline and alternative scenario projections. The Tealbook provides the point forecast and  70\%  intervals for the baseline. To construct the baseline distribution,  we take the point forecast as the median and extremes of the 70\% confidence interval as the 15th and 85th percentiles, respectively. We fit a skew-t with 50 degrees of freedom to these percentiles-- the choice of degrees of freedom allows some modest tail-weight beyond normal but effectively represents a \lq\lq close to normal" distribution.

\begin{table}[hb!]\caption{Tealbook Baseline and Scenarios}\label{tab:TBbaseline} \footnotesize
\setlength{\tabcolsep}{0\textwidth} \centering \medskip
\begin{tabular}{
L{.01\textwidth} L{.02\textwidth} L{.275\textwidth}R{.06\textwidth}R{.06\textwidth}R{.06\textwidth}
C{.015\textwidth}|C{.015\textwidth}
L{.02\textwidth} L{.275\textwidth}R{.06\textwidth}R{.06\textwidth}R{.06\textwidth}R{.01\textwidth}} 
&\multicolumn{5}{c}{\small \sc December 2007} &\multicolumn{2}{c}{ }& \multicolumn{5}{c}{\small \sc December 2018} \\ \hline\hline
&\blu $j$ &   \blu Scenario $\scn$&   \blu P50 &   \blu P15 &   \blu P85     &&&   \blu $j$   &   \blu Scenario $\scn$  &   \blu P50 &   \blu P15 &   \blu P85 &\\ \hline
&\blu 0   &   Baseline    &   1.3    &   0.1    &   2.5    &&&  \blu 0   &   Baseline    &   2.4 &   1.2 &   3.9 &\\
&\blu 1   &   Greater housing correction  &   1.0    &       &       &&& \blu 1   &   Financial-based recession   &	$-$0.7    &       &       &\\
&\blu 2   &   Credit crunch   &	$-$0.4   &       &       &&&    \blu 2   &   Stronger supply side    &   3.1 &       &       &\\
&\blu 3   &   Stronger domestic demand    &   1.7 &       &      &&&     \blu 3   &   Greater interest rate sensitivity  &  1.5  &       &       &\\
&\blu 4   &   With better export performance  &   1.9 &       &       &&&   \blu 4   &    Foreign slowdown   &   1.6 &       &       &\\
&\blu 5   &   Greater cost pressure   &   1.2 &       &       &&&   \blu    &      &    &       &       &\\
&\blu 6   &   Market-based federal funds rate &   1.6    &       &       &&&       &   &    &       &       &\\ \hline
\end{tabular}
\begin{tabular}{p{\textwidth}} \scriptsize
{\em December 2007 TB}: The baseline projection for 2008 is on page I-21, while the alternative scenarios are on page I-17---the scenarios for 2008 are obtained by averaging the values for 2008:H1 and 2008:H2.
{\em December 2018 TB}: The baseline projection for 2019 is on page 88, while the alternative scenarios are on page 84.     
\end{tabular}
\end{table}

\begin{figure}[ht!]\caption{Reference and baseline with scenario point forecasts} \label{fig:refandbaselinepdfcdf}
\centering \setlength{\tabcolsep}{.0\textwidth} \medskip

\begin{tabular}{C{.5\textwidth}C{.5\textwidth}}
\footnotesize 2007 & \footnotesize 2018 \\
\includegraphics[width=.48\textwidth]{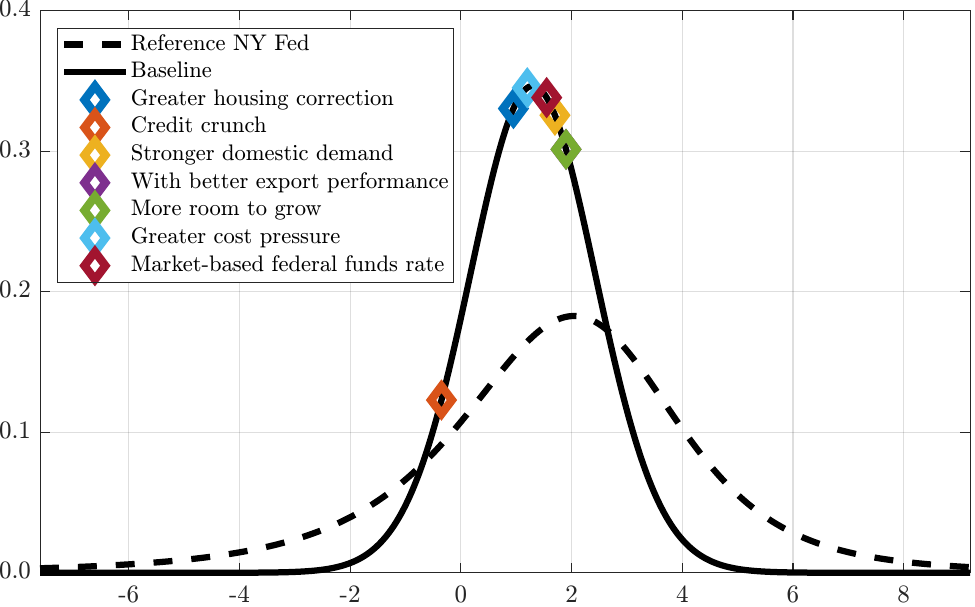}& 
\includegraphics[width=.48\textwidth]{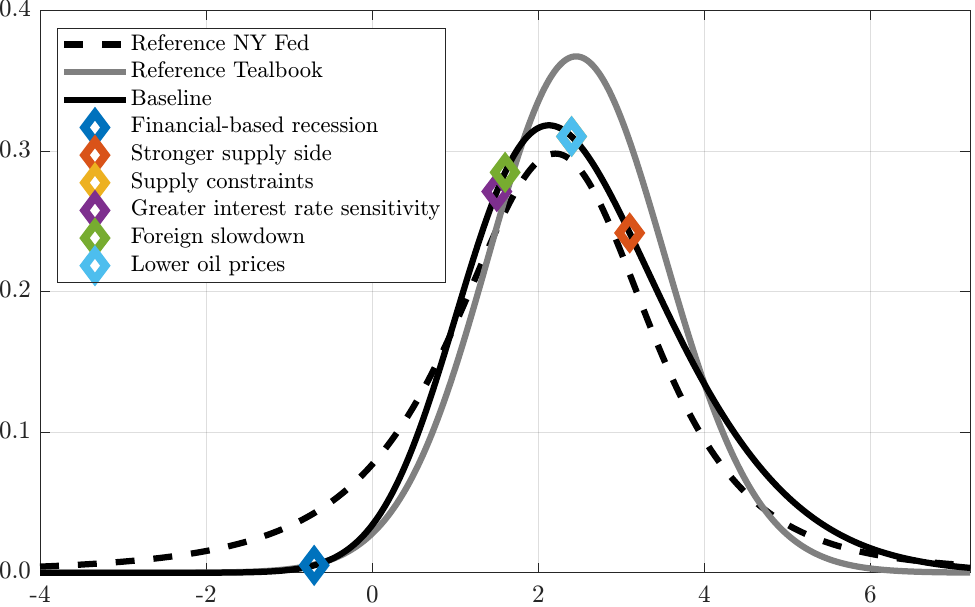} \\
\end{tabular}
\begin{tabular}{p{\textwidth}} \scriptsize
Solid black line is baseline p.d.f, and dashed black line is reference p.d.f. estimated from the NY Fed Outlook-at-Risk; solid gray line is the reference estimated from the Tealbook Time-Varying Macroeconomic Risk. Diamonds show point forecasts from   scenarios.
\end{tabular}
\end{figure}

\begin{wraptable}{r}{.6\textwidth}\vspace{-1pt}\caption{Reference and baseline skew-t  parameters}\label{tab:refandbaselineparameters} 
\setlength{\tabcolsep}{0\textwidth} \small \centering 
\begin{tabular}{C{.05\textwidth}L{.13\textwidth}L{.085\textwidth}R{.07\textwidth}R{.07\textwidth}R{.07\textwidth}R{.07\textwidth}R{.01\textwidth}}\hline\hline
&Distribution    &   Type   &   	\blu lc	&	\blu sc	&	\blu sk      &	\blu df	&\\\hline
\multirow{3}{*}{\rotatebox{90}{\blu  \ 2007}}
&&&&&&&\\[-10pt]
&Reference   &   NY Fed	&	2.7	&	2.2	&	$-$0.5	&	3.4	&\\[1pt]
&Baseline   &   	    &	1.3	&	1.1	&	0.0	&	50.0	&\\[-10pt] 
&&&&&&&\\\hline
\multirow{4}{*}{\rotatebox{90}{\blu \  2018}}
&&&&&&&\\[-10pt]
&Reference   &    NY Fed    &	2.5	&	1.3	&	$-$0.3	&	3.0	&\\[1pt]
&Reference   &    Tealbook	&	2.1	&	1.1	&	0.5	&	50.0	&\\[1pt]
&Baseline   &   	        &	1.2	&	1.9	&	2.1	&	50.0	\\[-10pt] 
&&&&&&&\\\hline
\end{tabular}
\begin{tabular}{p{.555\textwidth}}\scriptsize
The skew-t parameters are those for location (lc), scale (sc), skewness (sk) and degrees of freedom (df). 
\end{tabular}\vspace{-1pt} 
\end{wraptable}

Table~\ref{tab:refandbaselineparameters} shows parameters of the baseline and reference skew-t distributions; Figure~\ref{fig:refandbaselinepdfcdf} shows the p.d.f.s. The baseline is much more precise than the NY Fed reference, with a lower scale and higher degrees of freedom. This raises questions for economic forecasting and policy design. In a world with a known \lq\lq true model" of the economy, the baseline and  reference would be identical; as they differ in practice, interpreting the scenarios is the challenge.  
By comparison, the baseline is roughly as precise as the Tealbook reference, the latter having 50 degrees of freedom as a result of the optimization process in fitting the skew-t.  This suggests that the Tealbook reference may underestimate risk. 
As Federal Reserve Chair Jerome Powell said during the Press Conference following the January 2025 FOMC meeting, we should not be surprised as {\em \blu \lq\lq it is human nature, apparently, to underestimate [...] how fat the tails are.[...] We think of things in a normal distribution. And in the economy, it’s not a normal distribution.''}  

Given reference and baseline distributions, our analysis proceeds based on Monte Carlo sampling from the reference. In developments below, the MC sample size is $10^6$ and resulting MC analysis summaries stable and robust across reanalyses with such a large sample size.

%

\subsection{Scenarios\label{sec:GDPscenariodistributions}}

TB scenarios in Table~\ref{tab:TBbaseline} provide only point forecasts. In the Dec. 2007 (2018) TB, we have 6 (4) alternative scenarios.\footnote{We exclude  \lq\lq More room to grow" scenario in 2007, and both  \lq\lq Supply constraints" and \lq\lq Lower oil prices" scenarios in 2018. These had GDP point forecasts identical to either the baseline or another scenario. If included in our synthesis, they would receive equal weight with either the baseline or one other scenario.
} 
Figure~\ref{fig:refandbaselinepdfcdf} shows scenarios on the baseline, indicating their concentration around the baseline median but with some indication of downside risk in the left tail. 

Our first analysis treats the scenario point forecasts as medians\footnote{TB and other point forecasts might alternatively be treated as modes of scenario distributions.  Our analysis can address that, based on new theoretical results (not reported here) showing why and how entropic tilting can be applied when point forecasts are modes. 
We use medians, however, based on fundamental concern for deeper representation of the probability distributions of scenarios, and embedding in more detailed analyses with multiple percentiles.}
of the $p_j(y)$, and 
the  ET construction maps the baseline to each scenario p.d.f. constrained to its specified median only; see the P50s in Table~\ref{tab:TBbaseline}.
While the TB provides no measures of uncertainty around the scenario projections, such information could be useful and available in other applications. Section~\ref {sec:GDPsynth3P} explores analyses with P15 and P85 constructed for each scenario. 

As discussed in Section~\ref{sec:scenariomixturesBPS}, we augment the scenario set with
a \safehaven located at the center of the scenarios while being relatively over-dispersed. Since the scenario information here is restricted to the median point forecasts, we first construct $p_j(y)$, $j=1,\ldots,J$, using ET as in Section~\ref{sec:scenarioET}, then use the implied percentiles to define those of 
the \safehavennospace. Specifically, the \safehaven has
$\text{P50}_{\text{B}}=\text{\small median}_{j=1,\ldots, J}\text{P50}_j$, $\text{P15}_{\text{B}}=\min_{j=1,\ldots, J}\text{P15}_j$, and $\text{P85}_{\text{B}}=\max_{j=1,\ldots, J}\text{P85}_j$, respectively.  
Numerical details are in Table~\ref{tab:SynthMedianonly}.

\begin{figure}[ht!]\caption{Examples of Tilted Distributions of Alternative Scenarios \\ \small December 2007 Tealbook}\label{fig:ETscenariosGDPexample}
\centering
\setlength{\tabcolsep}{.0\textwidth}
\begin{tabular}{C{.5\textwidth}C{.5\textwidth}}
\footnotesize $\scn_1$: Greater housing correction & \footnotesize $\scn_2$: Credit crunch\\
\includegraphics[width=.48\textwidth]{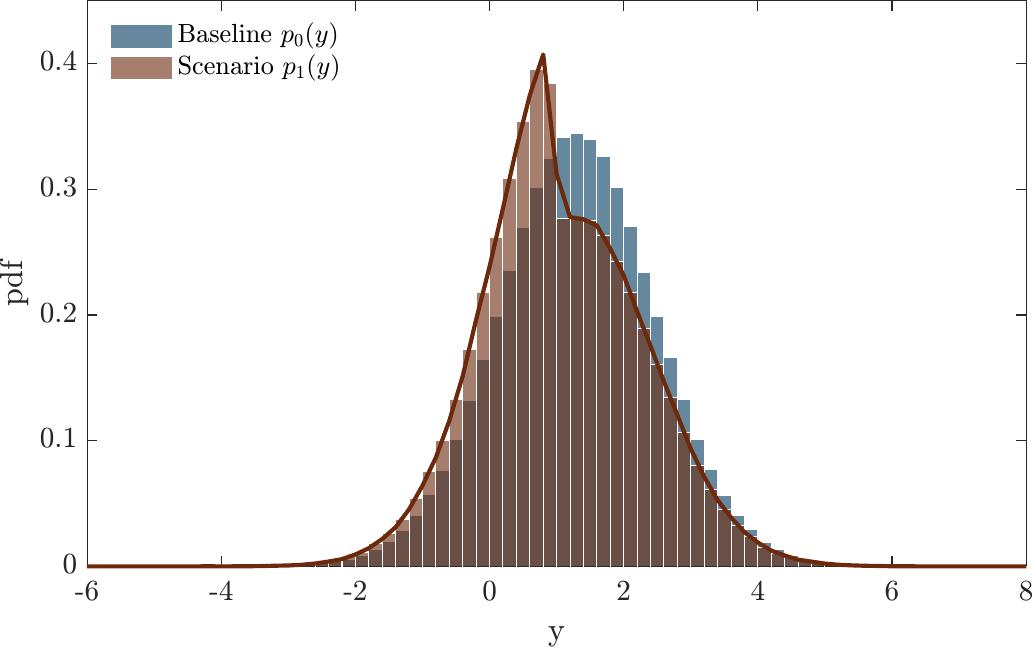}&
\includegraphics[width=.48\textwidth]{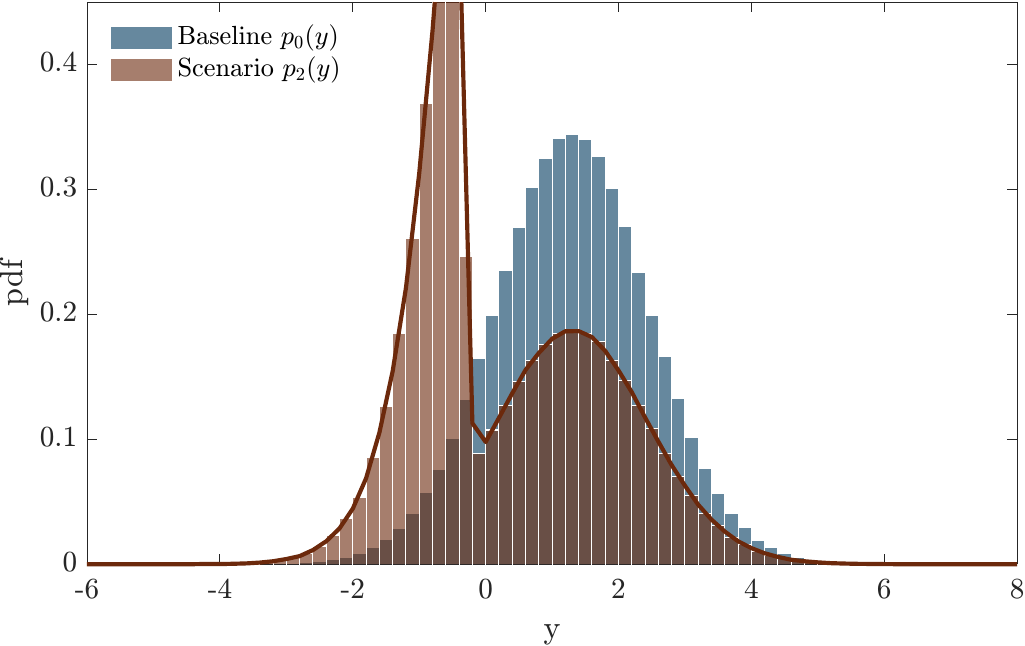}\\
\footnotesize $\scn_5$: Greater cost pressure & \footnotesize $\scn_7$: \Safehaven\\
\includegraphics[width=.48\textwidth]{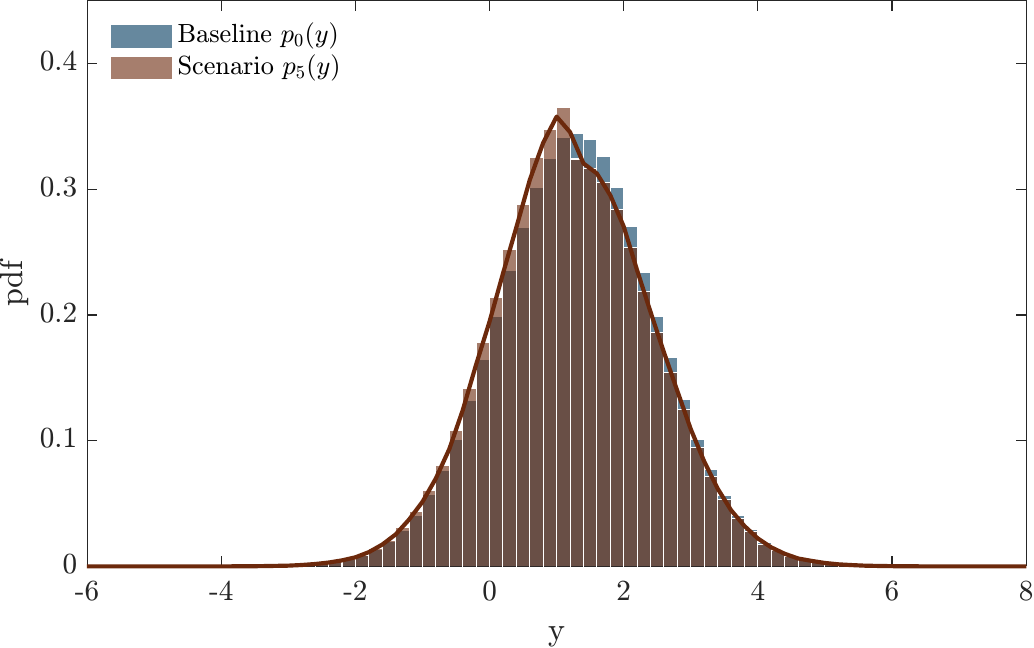}&
\includegraphics[width=.48\textwidth]{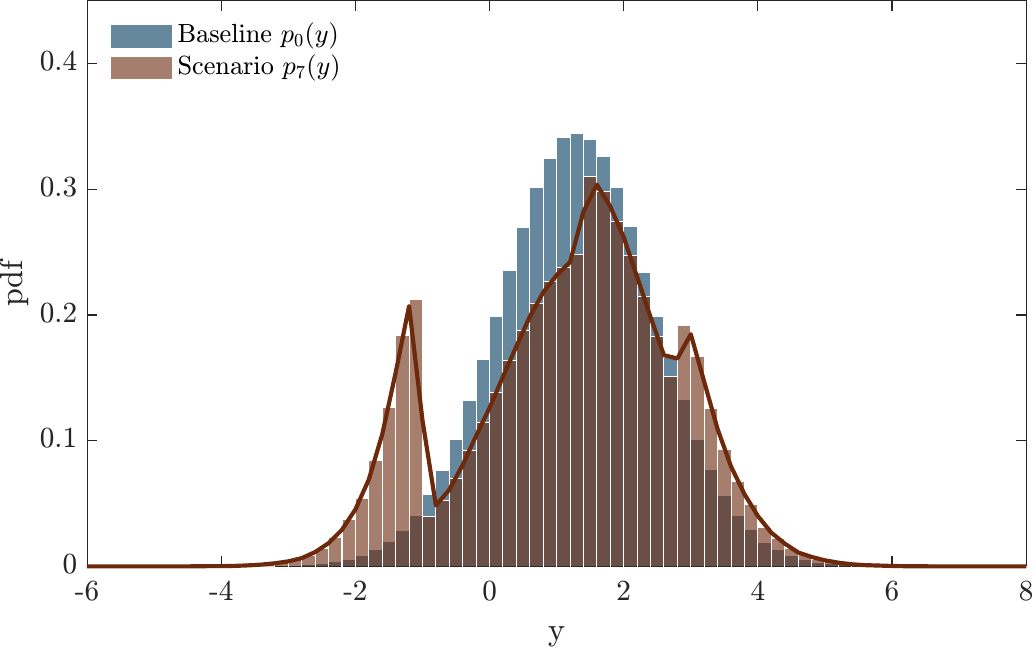}\\
\end{tabular} 
\end{figure} 

Figure~\ref{fig:ETscenariosGDPexample} shows the resulting $p_j(y)$ for four of the 2007 TB scenarios. 
Table~\ref{tab:SynthMedianonly}  shows corresponding 
ET--based ESS measures.  When a scenario is close to the baseline (e.g., $\scn_1$ and $\scn_5$), the tilted distribution remains close and slightly asymmetric; otherwise, the tilted distribution can exhibit skewness and multimodality due to the mismatch between the scenario forecasts and baseline.
The emergence of interesting shapes and multimodality in scenario distributions indicates the hypothesized state of the economy in the scenario is in regions poorly supported by the baseline. This is also related to the concept of modest policy intervention of~\cite{LeeperZha}, i.e. that we can analyze policy effects using the baseline as long as entertained policy interventions are small enough that economic agents would not change their behavior in response to the intervention. Related considerations are those of formally down-weighting extreme conditional assumptions in conditional forecasting~\citep[e.g.][sect.~3]{AntolinDiazPetrellaRubioRamirez2021,ChernisTallmanKoopWest2024}.

%

\subsection{Scenario Synthesis based on Medians\label{sec:GDPsynthmedianonly}}

Table~\ref{tab:SynthMedianonly} shows optimal scenario mixture weights and other summaries from analyses. 
There is strong concordance between the mixture synthesis and the reference in the case of the 2018 TB (EMR=0.48), while the concordance is weaker in the 2007 Tealbook (EMR=0.43). Figure~\ref{fig:optscenmixrefmixrefsafehaven} provides insights via 
comparison of p.d.f.s.   For TB 2008, the mixture synthesis is  light-tailed relative to the reference, as all scenarios lack probability on downside GDP ranges that the reference meaningfully supports. For the 2018 example, the scenario mixture supports positive GDP values-- partly as the baseline is already right skewed-- but assigns  relatively limited support to negative GDP growth.

\begin{table}[t!]\caption{Synthesis based on P50 information: NY Fed reference}\label{tab:SynthMedianonly}
\small\centering \setlength{\tabcolsep}{.0\textwidth} \medskip

\begin{tabular}{C{.04\textwidth}L{.025\textwidth}L{.3\textwidth}R{.07\textwidth}R{.07\textwidth}R{.07\textwidth}R{.08\textwidth}R{.08\textwidth}R{.08\textwidth}R{.08\textwidth}R{.08\textwidth}R{.02\textwidth}} \hline\hline
&&&&&&&&&&\\[-10pt]
\blu TB &\blu $j$ & \blu  Scenario $\scn$  & \blu P15$_j$ & \ \blu P50$_j$ & \ \blu P85$_j$   & \blu ET$_j$\%   & \blu  IS$_j$\%  & \blu $\pi_j^*$ & \blu $\alh_j$  & \blu $\als_j$ \\[-10pt]
&&&&&&&&&&\\\hline
&&&&&&&&&&\\[-10pt]
\multirow{10}{*}{\rotatebox{90}{\blu  Dec. 2007}}
&\blu 0  &  Baseline                         &   0.1    &   1.3     &  2.5  &   100.0   &    62.6    &  0.41  &  0.31   &   0.27    &\\[1pt]
&\blu 1  &  Greater housing correction       &  $-$0.1  &   0.9     &  2.3  &  94.3 &   57.4    &  0.40  &  0.00   &   0.02    &\\[1pt]
&\blu 2  &  Credit crunch                    &  $-$1.0  &  $-$0.4   &  2.0  &  28.7 &   30.9    &  0.36  &  0.08   &   0.08    &\\[1pt]
&\blu 3  &  Stronger domestic demand         &   0.3    &   1.7     &  2.7  &  92.6 &   65.4    &  0.42  &  0.00   &   0.04    &\\[1pt]
&\blu 4  &  Better export performance        &   0.4    &   1.9     &  2.9  &  84.3 &   65.2    &  0.42  &  0.31   &   0.27    &\\[1pt]
&\blu 5  &  Greater cost pressure            &   0.0    &   1.2     &  2.5  &  99.5 &   61.3    &  0.41  &  0.00   &   0.02    &\\[1pt]
&\blu 6  &  Market-based Fed Funds rate      &   0.2    &   1.6     &  2.6  &  97.1 &   64.8    &  0.41  &  0.00   &   0.03    &\\[1pt]
&\blu 7  &  \Safehaven                       &  $-$1.0  &   1.4     &  2.9  &  56.4 &   67.2    &  0.43  &  0.31   &   0.27    &\\[-10pt]
&&&&&&&&&&\\\cdashline{2-11}
&&&&&&&&&&\\[-10pt]
&       &  \blu $f(\y|\alh)$                     &  $-$0.2  &   1.4  &  2.7  &     &   71.5    &  0.43  &  &       &\\[1pt]
&       &  \blu $f(\y|\als)$                     &  $-$0.2  &   1.4  &  2.7  &     &   71.2    &  0.43  &  &       &\\[-10pt]     
&&&&&&&&&&\\\hline
&&&&&&&&&&\\[-10pt]
\multirow{8}{*}{\rotatebox{90}{\blu  Dec. 2018}}
&\blu 0  &  Baseline                           &   1.2  &   2.4  &  3.9  &   100.0  &   88.5  &  0.47  &  0.74  &    0.64   &\\[1pt]
&\blu 1  &  Financial-based recession          &  $-$1.0  &  $-$0.6  &  3.1  &   0.6  &   8.4  &  0.35  &  0.04  &    0.04   &\\[1pt]
&\blu 2  &  Stronger supply side               &   1.4  &   3.1  &  4.4  &  84.6  &  67.5  &  0.45  &  0.00  &    0.04   &\\[1pt]
&\blu 3  &  Greater interest rate sensitivity  &   0.7  &   1.5  &  3.4  &  69.4  & 70.3    &  0.45  &  0.13  &    0.11   &\\[1pt]
&\blu 4  &  Foreign slowdown                   &   0.8  &   1.6  &  3.5  &  75.3  & 74.5    &  0.46  &  0.02  &    0.10   &\\[1pt]
&\blu 5  &  \Safehaven                     &  $-$1.0  &   1.6  &  4.4  &   2.1  &   37.9    &  0.43  &  0.07  &  0.08     &\\[-10pt]
&&&&&&&&&&\\\cdashline{2-11}
&&&&&&&&&&\\[-10pt]
&       &  \blu $f(\y|\alh)$                     &  0.9  &   2.2  &  3.8  &     &   91.4    &  0.48  &  &       &\\[1pt]
&       &  \blu $f(\y|\als)$                     &  0.9  &   2.2  &  3.8  &     &   90.9    &  0.48  &  &       &\\[-10pt]
&&&&&&&&&&\\\hline
\end{tabular}
\begin{tabular}{p{\textwidth}} \scriptsize
P15 and P85 for $\scn_j$, $j=1:J-1$, are for the ET--based $p_j(y)$; the other percentiles are inputs to the analysis. ET$_j$ is the ET-based ESS of $p_j(y)$ relative to the baseline $p_0(y)$ while IS$_j$ denotes that for 
ET--IS implied relative to the reference $p(y).$  $\pi_j^*$ is the EMR of the reference and scenario $j$ alone. $\alpha_j^*$ and $\widetilde\alpha^*$
are the probabilities on $\scnj$ in the optimal 
mixture synthesis in analyses with and without the \safehaven scenario, respectively.
The optimization constraints $\alpha_0\ge \alpha_j, j=\seq 1J,$ apply in both cases. 
\end{tabular} \smallskip
\end{table}

\begin{figure}[htb!]\caption{p.d.f. and c.d.f of scenario synthesis and NY Fed reference} \label{fig:optscenmixrefmixrefsafehaven}
\centering \setlength{\tabcolsep}{.0\textwidth}
\begin{tabular}{C{.5\textwidth}C{.5\textwidth}}
\scriptsize \sc December 2007 &  \scriptsize \sc December 2018  \\
\includegraphics[width=.49\textwidth]{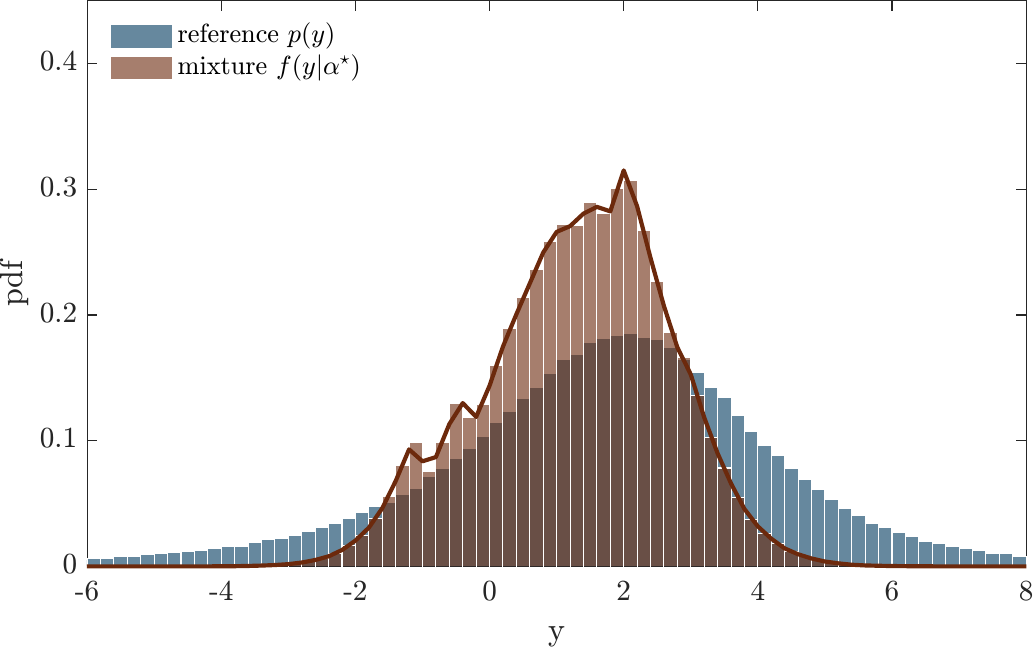} &
\includegraphics[width=.49\textwidth]{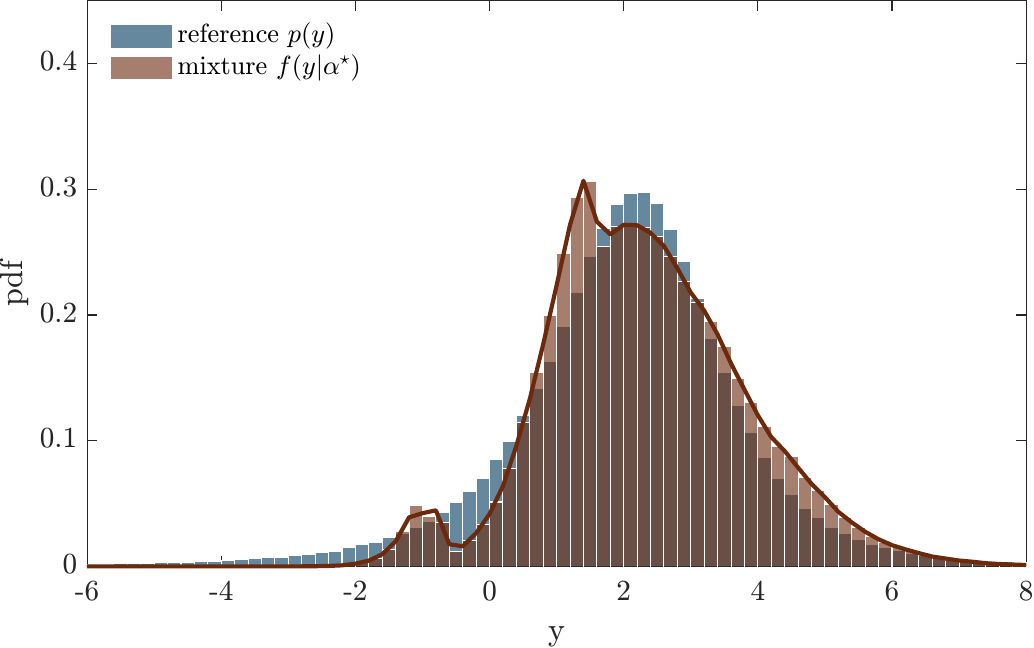}  \\
\includegraphics[width=.49\textwidth]{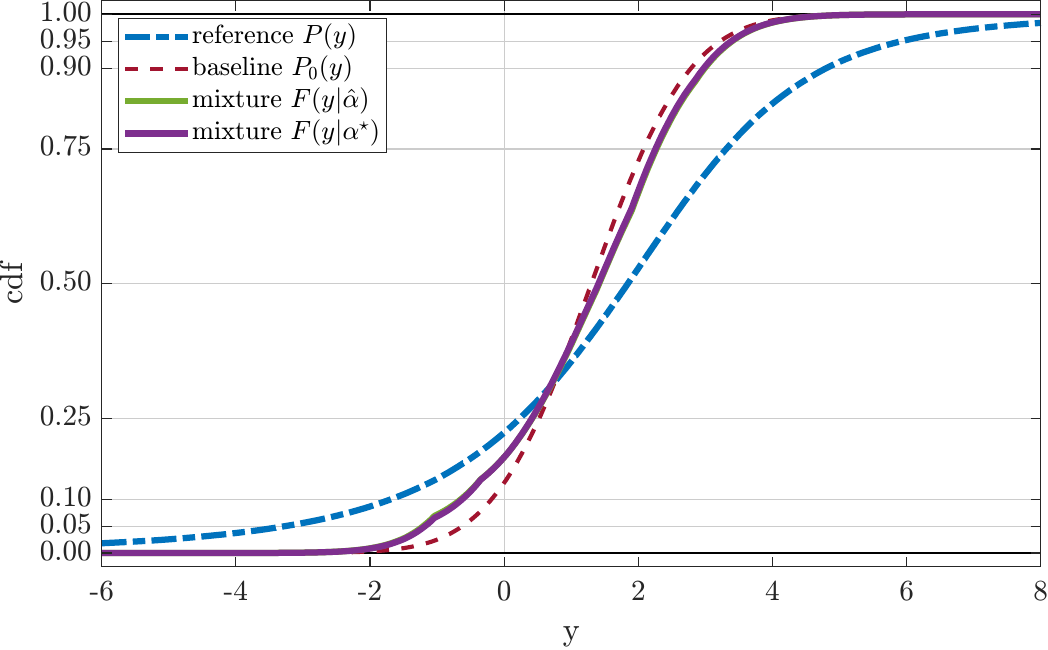}  &
\includegraphics[width=.49\textwidth]{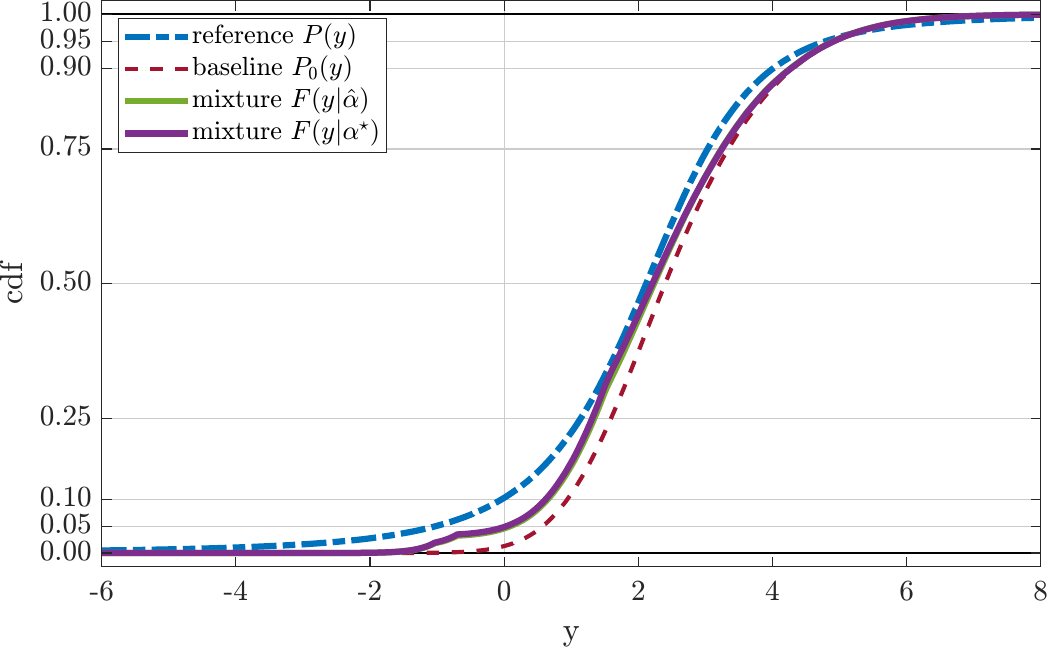}  \\
\end{tabular}
\begin{tabular}{p{\textwidth}} \scriptsize
{\em Upper:} p.d.f.s of the reference and scenario synthesis.
{\em Lower:}  c.d.f.s of the reference, baseline and scenario synthesis including results synthesis c.d.f.s based on both $\alh$ and $\als;$ note that the latter are effectively indistinguishable.  
\end{tabular}
\end{figure}

In the 2007 TB analysis, the baseline, $\scn_4$ and  \safehaven are roughly equally weighted, followed by    $\scn_2$. The two more extreme scenarios $\scn_2$ and $\scn_4$ get weight as they help to capture the spread of the reference, while the other scenarios get small weights-- they sit in the center of the reference and are very similar to the baseline, as shown by the ET ESS measures.  In contrast, in the 2018 TB example the reference is more precise and the baseline is heavier weighted than other scenarios. Here the EMR of the baseline alone is quite high, so the alternative scenarios add small but limited value in contributing to approximating the reference.


A key feature of analysis is that $100{-}\textrm{ESS}$  for the synthesis is an absolute measure of scenario set incompleteness relative to the reference. In TB 2007, the ESS of $f(\y|\als)$ is about 71-72\%; we can say that the scenario set is about 28-29\% incomplete. In contrast, in TB 2018 the ESS of $f(\y|\als)$ is about 91\%, which indicates that the TB scenarios much more adequately represent the risk and uncertainty in the economy as defined by the reference than in 2007. A hint of the inability of TB 2007 scenarios to properly capture the risks in the reference comes also from the substantial weight on the \safehavennospace; in contrast, in TB 2018, the \safehaven receives low weight. 
As a general point looking ahead, a \safehaven p.d.f. that is over-dispersed has general benefits, but in application other choices are possible and may be preferred.  These examples highlight this, indicating that scenarios reflecting increased support in the upper and lower tails of the-- anticipated-- reference distribution are most relevant.   Future applications might address this.

\begin{table}[htbp!]\caption{Synthesis based on P50 information: TB 2018, Tealbook reference}\label{tab:Synth2018V2} \centering
\small\setlength{\tabcolsep}{.0\textwidth} \medskip
\begin{tabular}{C{.01\textwidth}L{.025\textwidth}L{.3\textwidth}R{.07\textwidth}R{.07\textwidth}R{.07\textwidth}R{.08\textwidth}R{.08\textwidth}R{.08\textwidth}R{.08\textwidth}R{.08\textwidth}R{.02\textwidth}} \hline\hline
&&&&&&&&&&\\[-10pt]
 &\blu $j$ & \blu  Scenario $\scn$  & \blu P15$_j$ & \ \blu P50$_j$ & \ \blu P85$_j$   & \blu ET$_j$\%   & \blu  IS$_j$\%  & \blu $\pi_j^*$ & \blu $\widehat{\alpha}_j$  & \blu $\alpha_j^*$ \\[-10pt]
&&&&&&&&&&\\\hline
&&&&&&&&&&\\[-10pt]
&\blu 0  &  Baseline                           &       1.2  &       2.4    &    3.9    &    100.0   &    77.7    &    0.49     &    1.00   &    0.89    &\\[1pt]
&\blu 1  &   Financial-based recession         &    $-$1.0  &    $-$0.6    &    3.1    &     0.6    &     0.8    &    0.33     &    0.00   &    0.01    &\\[1pt]
&\blu 2  &  Stronger supply side               &       1.4  &       3.1    &    4.4    &    84.7    &    51.0    &    0.47     &    0.00   &    0.05    &\\[1pt]
&\blu 3  &  Greater interest rate sensitivity  &       0.7  &       1.5    &    3.4    &    69.6    &    56.3    &    0.44     &    0.00   &    0.02    &\\[1pt]
&\blu 4  &  Foreign slowdown                   &       0.8  &       1.6    &    3.5    &    75.4    &    61.3    &    0.45     &    0.00   &    0.02    &\\[1pt]
&\blu 5  &  \Safehaven                        &     $-$1.0  &       1.6    &    4.4    &     2.1    &     3.2    &    0.40     &    0.00   &    0.01    &\\[-10pt]
&&&&&&&&&&\\\hdashline
&&&&&&&&&&\\[-10pt]
&       &  \blu $f(\y|\alh)$                     &  1.2  &   2.4  &  3.9  &     &   77.7    &  0.49  &  &       &\\[1pt]
&       &  \blu $f(\y|\als)$                     &  1.2  &   2.4  &  3.9  &     &   76.1    &  0.49  &  &       &\\[-10pt]
&&&&&&&&&&\\\hline
\end{tabular}
\begin{tabular}{p{.96\textwidth}} \scriptsize
Details as in Table~\ref{tab:SynthMedianonly}. 
\end{tabular} \smallskip
\end{table}

\FloatBarrier 

Table~\ref{tab:Synth2018V2} summarizes 2018 analysis using the reference distribution based on percentiles from the Tealbook Time-Varying Macroeconomic Risk to compare with the NY Fed-based details above.  Again the synthesis here uses only the scenario medians.   In this case, the baseline is as precise as the reference and has similar tail-weight in terms of the skew-t degrees of freedom. The main point, however, is that the baseline dominates the scenarios, indicating that the hypothesized median shifts they represent add little to no value in predictive discrimination relative to the reference.

On modeling strategy, consider an example of 
\lq\lq normal" economics times as represented by 2018. In such settings, 
(i) the reference can be expected to be relatively light-tailed, (ii) scenarios can be expected to be modest in terms of varying from \safehavennospace, and (iii) the resulting scenario synthesis will be be close to the reference. There is limited scenario set incompleteness and the \safehaven will  play a limited role.  Contrast this with periods of higher uncertainty, such as in the 2007 context here where the reference distribution should have appropriately fatter tails. Then the synthesis of the baseline and the scenarios can substantially under-represent the reference unless the scenario set includes more extreme considerations.  This mandates admitting extreme scenario considerations as a rational response to increased uncertainty in very uncertain economic times.

\subsection{Scenario Synthesis based on P15, P50 and P85\label{sec:GDPsynth3P}}
As earlier noted, the methodology admits specification of multiple features of the scenarios so long as they can be represented as expectations under implicit scenario distributions.  The case of multiple percentiles is practically key, and we visit this setting with summaries of further analysis in the TB 2007 context.

\begin{figure}[htbt!]\caption{p.d.f. and c.d.f of scenario synthesis and NY Fed reference}\label{fig:Synth2007P3}
\small\centering Tilted scenario distributions obtained using three percentiles --- December 2007 Tealbook \\
\setlength{\tabcolsep}{.0\textwidth} \medskip
\begin{tabular}{C{.5\textwidth}C{.5\textwidth}}
\includegraphics[width=.49\textwidth]{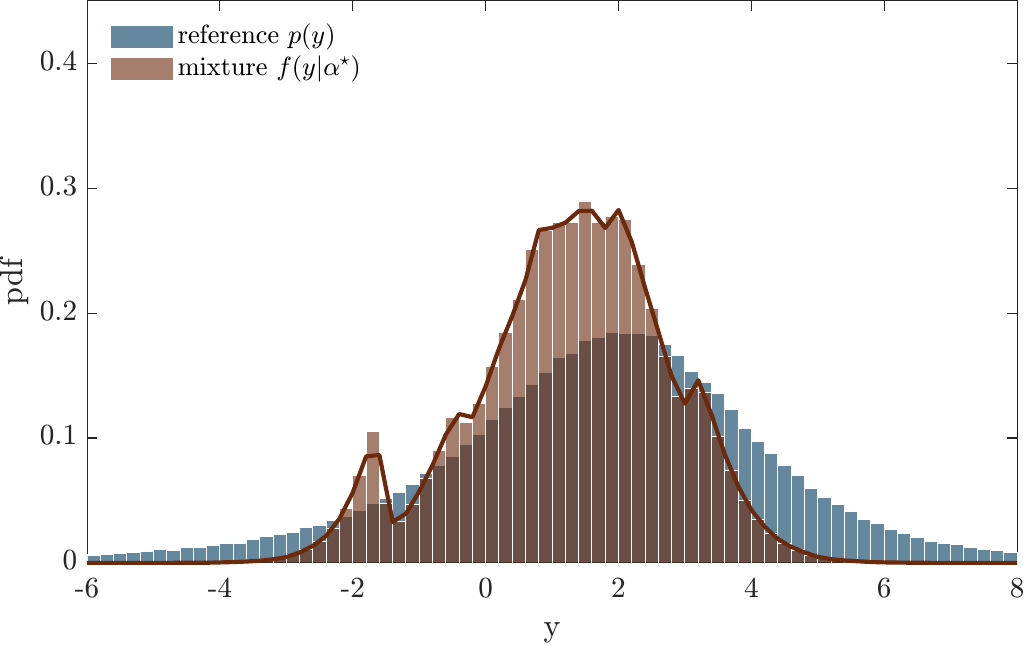} &                  
\includegraphics[width=.49\textwidth]{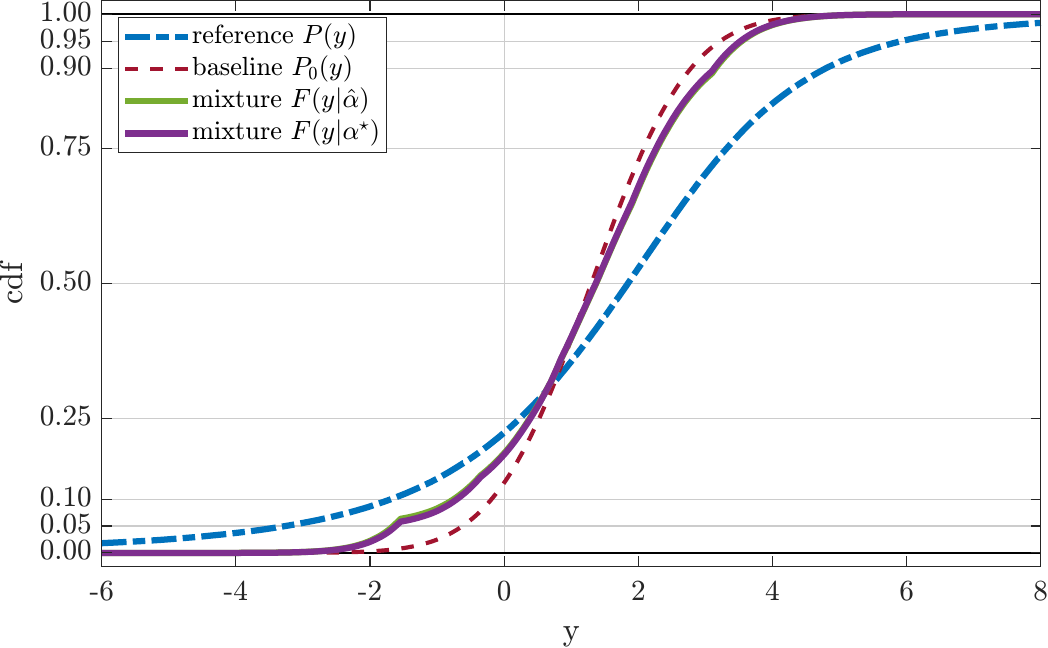}\\
\end{tabular}
\begin{tabular}{p{\textwidth}} \scriptsize
{\em Left:} p.d.f. of the reference and scenario synthesis. {\em Right:}  c.d.f. of the reference, baseline and scenario synthesis, the latter comparing results using $\alh$ and $\als$ for the synthesis; again, the latter are effectively indistinguishable.
\end{tabular}
\end{figure}

Figure~\ref{fig:Synth2007P3} and Table~\ref{tab:Synth2007P3} report the results when scenario p.d.f.s are  tilted versions of the baseline that match the scenario-specific P15, P50, and P85.  Here the scenario P15 and P85 are computed assuming distances of tail percentiles from median agree with the baseline: $\text{P15}_j=\text{P50}_j-(\text{P50}_0-\text{P15}_0)$ and $\text{P85}_j=\text{P50}_j+(\text{P85}_0-\text{P50}_0)$. The P15 and P85 for the scenarios in Table~\ref{tab:Synth2007P3} are similar to those in Table~\ref{tab:SynthMedianonly} when the scenario is close to the baseline; they are naturally more discordant with increasing departure of the scenario percentiles from those of  baseline (e.g.,  $\scn_2$). 
Then, optimal weights and EMR result are quite similar to those in Table~\ref{tab:SynthMedianonly}. Other specifications of the P15 and P85 uncertainties-- specifications that are founded in scenario considerations-- may be quite different than the synthetic choices here, of course, and can be expected to lead to different results. A main point is that the methodology is open to-- and trivially applied to-- scenario specifications in terms of multiple percentiles,  with negligible analytic and computational burden. Such specifications are increasingly common in application, and to be encouraged in policy research moving forward. 

\begin{table}[htbp!]\caption{Synthesis based on P15, P50 and P85 information: TB 2007, NY Fed reference}\label{tab:Synth2007P3} \centering
\small\setlength{\tabcolsep}{.0\textwidth} \medskip
\begin{tabular}{C{.01\textwidth}L{.025\textwidth}L{.3\textwidth}R{.07\textwidth}R{.07\textwidth}R{.07\textwidth}R{.08\textwidth}R{.08\textwidth}R{.08\textwidth}R{.08\textwidth}R{.08\textwidth}R{.02\textwidth}} \hline\hline
&&&&&&&&&&\\[-10pt]
 &\blu $j$ & \blu  Scenario $\scn$  & \blu P15$_j$ & \ \blu P50$_j$ & \ \blu P85$_j$   & \blu ET$_j$\%   & \blu  IS$_j$\%  & \blu $\pi_j^*$ & \blu $\widehat{\alpha}_j$  & \blu $\alpha_j^*$ \\[-10pt]
&&&&&&&&&&\\\hline
&&&&&&&&&&\\[-10pt]
&\blu 0  &  Baseline                         &   0.1  &   1.3  &  2.5  &   100.0    &   62.6    &  0.41  &  0.30   &   0.26    &\\[1pt]
&\blu 1  &  Greater housing correction       &  $-$0.2  &   1.0  &  2.2  &  92.3    &   57.2    &  0.39  &  0.00   &   0.01    &\\[1pt]
&\blu 2  &  Credit crunch                    &  $-$1.5  &  $-$0.3  &  0.9  &  20.0  &   31.6    &  0.32  &  0.10   &   0.11    &\\[1pt]
&\blu 3  &  Stronger domestic demand         &   0.5    &   1.7     &  2.9  &  90.1 &   65.9    &  0.42  &  0.00   &   0.07    &\\[1pt]
&\blu 4  &  Better export performance        &   0.7    &   1.9  &  3.1  &  79.3   &   66.1   &  0.42  &  0.30   &   0.26    &\\[1pt]
&\blu 5  &  Greater cost pressure            &   0.0  &   1.2  &  2.4  &  99.4  &   61.2    &  0.40  &  0.00   &   0.01    &\\[1pt]
&\blu 6  &  Market-based Fed Funds rate      &   0.4  &   1.6  &  2.8  &  96.0  &   65.0    &  0.42  &  0.00   &   0.03    &\\[1pt]
&\blu 7  &  \safehaven                       &  $-$1.5  &   1.4  &  3.1  &  27.7 &   62.4    &  0.43  &  0.30   &  0.26         &\\[-10pt]
&&&&&&&&&&\\\cdashline{2-11}
&&&&&&&&&&\\[-10pt]
&       &  \blu $f(\y|\alh)$                  &  $-$0.3  &   1.4  &  2.8  &     &   73.0    &  0.44  &  &       &\\[1pt]
&       &  \blu $f(\y|\als)$                  &  $-$0.3  &   1.4  &  2.8  &     &   72.7    &  0.44  &  &       &\\[-10pt]
&&&&&&&&&&\\\hline
\end{tabular}
\begin{tabular}{p{.97\textwidth}} \scriptsize
Details as in Table~\ref{tab:SynthMedianonly}. 
\end{tabular} \smallskip
\end{table}

\section{Distributional Forecasts and Judgment  \label{sec:judgmentaltilting}}
 
At a general level, our focus is on use and reconciliation of information from statistical models and judgmental sources. Analysis is directional in that the specific goals are to assess judgmentally derived scenarios against a statistical reference distribution. In the broader context, the reverse is also of interest; that is, investigation of how a statistical forecast distribution may be \lq\lq tilted" in a direction deemed important from a judgmental point of view. The latter can often be proxy for information external to that underlying the statistical model.

A key context is that of unique, unexpected events and  shifts in the structure of the economy that go well beyond existing model structure and assumptions.  While structural economic models may provide a formal basis for longer-term adaptation, fully modeling the implications of regime shifts on the structure of the economy takes time. Short-term, judgmental adjustments can be most valuable for real-time decision making. Indeed, judgment plays a dominant role in decision making in other areas, such as among investment professionals in macroeconomic trading. In monetary policy settings, quantitative macroeconomic models provide a firmer basis, but undesirable policy recommendations from models are often attributed to persistent forecast errors. A \lq\lq good'' policy maker would intervene to input judgment to address this. 

Subjective information sources include surveys, market intelligence, and stress test outputs. Some comments on each are germane. 

{\em i. Surveys.} Perhaps the key example is the US Survey of Professional Forecasters (SPF) a well-established source of community-wide 
forecast information. SPF now collates probabilistic forecasts on predefined bins for outcomes ~\citep{Croushore1993,DelNegro2023}.   
Similar regular surveys are conducted by ECB, the Bank of England, and other institutions. 

{\em ii. Market Intelligence.} Large and detailed information sets are
commonly collected by central banks in order to inform policy makers on many dimensions of economic and financial market developments outside the scope of well-adopted structural macroeconomic models. Such models, aiming to
reduce complexity and lead to openness and interpretation in economic terms, inevitably lack the ability to reflect the full complexity of structural changes or nonlinear dynamics that become practically relevant in more unusual circumstances. More complex economic and financial market intelligence-- in the form of summary external forecast information and judgment-- can then add real value, if recognized and appropriately integrated with the model-based forecasts.  

{\em iii. Stress testing.}  Initially developed by the IMF in the 1990s
to assess financial system resilience, stress testing was later broadly 
adopted to assess macroeconomic risks from banking distress
following the Great Financial Crisis~\citep{Adrian2020}.
All major financial regulators now design and publish macroeconomic stress scenarios and assess financial stability relative to those scenarios. 
This typically includes scenarios of major downturns in macroeconomic aggregates such as real activity, inflation, and financial conditions.
Scenario design emphasizes extreme economic and financial circumstances; the resulting outputs can provide a basis for judgmental modification of forecasts from the established reference econometric models that are not designed or customized to quickly and easily address such circumstances. 
 
The overall question is that of intervention in the statistical model to incorporate such external information.  The concept is long recognized and much methodology exists and is used in other areas of forecasting, such as commercial and financial 
applications~(e.g.,~\citealp{WestHarrision.JASA.1986};
\citealp{WestHarrison.JoF.1989};
\citealp{black1991asset};
\citealp[][chap.~11]{WestHarrison.YellowBook2ndEdn.1997};
\citealp{West2023constrainedforecasting}). 
However, emphasizing and formalizing the question with respect to policy applications is highlighted and of renewed interest here. 

Beyond contextual connections with the main theme of the current paper, key technical features of our scenario synthesis methodology relate directly to these complementary interests.  From Section~\ref{sec:Methodology} and generalizing the notation there, each of the constructed scenario p.d.f.s has the form $p_j(\y) \propto w_j(\y) p(\y)$ based on the statistical reference p.d.f. $p(\y)$ and scenario-specific weight-- or tilting-- function $w_j(\y).$ The latter is 
$w_j(\y) \propto w_0(\y) \exp\{ \btau_j'\s_j(\y)\} $ involving: (i) the ET term 
$\exp\{ \btau_j'\s_j(\y)\}$ 
used to define $p_j(\cdot)$ using the baseline and partial scenario information;  and (ii) 
the baseline-reference IS weight function $w_0(\y) \propto p_0(\y)/p(\y).$  The Monte Carlo 
methodology uses the discrete versions over the reference random samples $\y^i,$ i.e., 
$w_j^i = w_j(\y^i)$ for $j=\seq 0J.$ 

The normalized scenario p.d.f.s are then $p_j(\y) = c_j w_j(\y) p(\y)$ where the $c_j$ are just normalizing constants. Thus, explicitly, the partial judgmental information that $\scnj$ encodes leads to a weighted modification of the statistical reference; on $\scnj$ alone, this can be regarded as the scenario-tilted version of the model-based forecast.   This indicates that the overall question of conditioning a model-based forecast on what may be quite distinct forms of judgmental information is intimately addressed within our framework.  It also follows that the BPS-justified scenario mixture
p.d.f. is $\fya = w(\y|\balpha) p(\y)$ where $w(\y|\balpha) = \sum_{j=\seq 0J} \alpha_j c_j w_j(\y)$. This is true for any $\balpha,$  not just the EMR-optimized value central to our scenario synthesis goals. In other contexts, such as the above settings of modifying the reference $p(\y)$ with judgment-based 
information summaries, this allows for context-specific specification of relative scenario weightings. Importantly, scenarios can address multiple aspects of the forecast distribution, including location shifts, scale and skewness perturbations-- both within any one scenario and with diversity across a scenario set. This may be particularly important to extension and evaluation of this approach in areas such as stress testing.

\section{Summary Comments\label{sec:conclusions}}

The formal assessment and integration of partial scenario information with statistical forecast distributions is of interest in a range of policymaking settings. 
Our approach has been motivated by the monetary policy process, where policy decisions are firmly rooted in macroeconomic forecasts that involve not only the baseline forecast, but also alternative risk scenarios. The methodology offers a concrete and straightforward approach to evaluating baseline and judgmental scenario assessments-- with their intuitive and easily communicated bases-- 
against more formal statistical density forecasts of risk.  Applications to the monetary policy process are highlighted, and offer a new frontier for practical yet rigorous policy making.

We believe the methodology will have broad appeal in other applications. For example, the IMF continuously monitors global financial stability, and publishes a formal biannual GaR-based global financial stability assessment in its Global Financial Stability Report. The statistical approach can be compared directly-- and in a quantitatively meaningful manner-- with the scenario-based risk assessment of the IMF's World Economic Outlook, published on the same schedule as the GFSR. 

The approach can also be readily applied to other areas of institutional risk management and contexts such as portfolio choice applications. In risk management, as in financial institution supervisory stress testing, both scenario-based approaches and more statistical approaches are commonly deployed. Our framework and methodology offers a novel way to evaluate and integrate those two avenues in a concrete fashion. In portfolio allocation decision making, the role of priors is fundamental and features commonly in allocation decisions, yet the bridge between intuitive scenario based approaches and statistical modeling of return forecasts has received little attention. Again our framework offers steps ahead in this regard.  Beyond these areas, there are also opportunities in commercial revenue and supply chain forecasting where ranges of forms of external/subjective information are often assessed in the context of formal models with a view to eventual decisions. 


Forms of scenario information that may feed-into new applications are information sets with more than a few candidate percentiles of forecast distributions under any assumed scenario. If a scenario is specified in terms of a larger number of percentiles, then analysis begins to approximate that given a fully-specified scenario p.d.f. $p_j(\y).$ This is certainly of methodological interest, and may represent applied interests in settings where the scenarios are effectively replaced by forecast distributions from alternative/competing models. This latter setting is closer to the existing setting of BPS where multiple predictive distributions are considered by a Bayesian decision maker, and define analogue information to condition an initial reference forecast distribution.  Some of the technical developments here are rather different-- and complementary-- to the general setting of BPS, but open up new questions for potential development and exploitation in forecast model comparison and synthesis. There are also questions of extension of the technical approach to address synthesis based on other forms of scenario information, e.g., point forecasts that are regarded as subjectively assessed modes or means rather than medians, and uncertainty in scenario information, e.g., percentiles provided with some notational $\pm$ uncertainties. The general ET framework in principle applies to such contexts, though details are to be developed for exploitation in any new applied context in which such scenario information sets arise.

We have noted that the methodology is applicable  with $\y$ in several dimensions.  Elements of $\y$ can include multiple economic indicators (real growth, inflation, unemployment, etc.,) as well as-- anchored at a current time period, such as the end of the current quarter-- multiple time periods ahead, such as the coming eight quarters.  Scenario information to define (uncertain) constraints on forecasts of the state of the macroeconomy over multiple future time periods can than generate a range of scenarios. Technically and computationally, the methodology here extends immediately.  We have experience with such extensions, and recognize questions that arise due to increasing dimension of $\y$.  In technical essentials, the main questions there are not new, but have to do with scalability of importance sampling methodology with dimension, and of its close technical 
ally entropic tilting with increasing dimension of the underlying Bayesian decision-analytic utility (a.k.a. score) functions. These questions are addressed in all applications of these general approaches, and will need to be addressed in context-- in specific applied settings of scenario synthesis.

\if0\blind{ 
\section*{Acknowledgments}
The authors thank colleagues at the Federal Reserve and the IMF-- including Gianni Amisano, Matthias Paustian, Sheheryar Malik, Jason Wu, and Pierre-Olivier Gourinchas-- for multiple discussions and feedback on the general area and a range of specific topics. We also thank Todd Clark of the Center for Financial Economics at Johns Hopkins University for his detailed, thoughtful comments on a first draft of our paper.  
The views expressed in this paper are those of the authors and do not necessarily reflect the views of the Federal Reserve System, the Federal Open Market Committee, or the International Monetary Fund, its Management, or its Executive Directors.  
 }\fi  
 
\FloatBarrier

\newpage 
\small
\bibliography{ScenarioSynthesis2025}
\bibliographystyle{chicago} 
\normalsize

\newpage
\begin{appendices} 
 
\section{Relating KL Divergences to EMR\label{app:moreEMRandKL}}

\subsection{Lower Bounds on EMR\label{app:KLboundEMR}} 

As noted in Section~\ref{sec:EMRandKL}, the bound $\emr\ge 1/[1+\exp\{\klpf\}]$ has empirical support in specific examples. This is, of course, not true in general, though seems suggested under certain, practically relevant conditions on $y\sim \py.$ The 
implied distribution of 
$\ky=\log\{\py/\fy\}$ has mean
$\tE[ \ky |\cH_p ]=\klpf \ge 0$ with equality only when $\ky=0$ for all $\y.$  The bound
is conjectured to hold when the distribution of $\ky$ has 
finite, positive mean, is unimodal with
$\tPr[\ky>0]>0.5,$ and has p.d.f. tail decay on 
$\ky<0$ no heavier than that on $\ky>0.$  
An exact proof for more restricted cases is available, as follows. 

Simplifying notation, the real-valued quantity $k$ replaces $\ky$. The focus is on 
$\emr = \tE_k[\pi(k)]$ where $\pi(k)=1/\{1+\exp(k)\}$ and   $k$ has some p.d.f. $g(k)$ with finite mean $m>0.$ Here 
 $\tE_k[\cdot]$ denotes expectation with respect to $k\sim g(k).$     
The following theory draws on standard results concerning scale mixtures of normals~\citep{AndrewsMallows1974,West1987}.

Suppose that $g(k)$ is continuous, symmetric and unimodal with mode and finite mean $m$. Then $g(k)$ is a normal scale mixture: 
$g(k)=\tE_v[ v^{-1}\phi\{v^{-1}(k-m)\} ]$ where $\phi(\cdot)$ is the standard normal p.d.f., $v$ is a random scale parameter 
 and $\tE_v[\cdot]$ denotes expectation with respect to its distribution.  
 
Then,  recognize that $\pi(k)=1/\{1+\exp(k)\}$ is the survival function of the standard univariate logistic distribution for real-valued $k.$   
The logistic distribution is also a normal scale mixture, so $\pi(k) = 1-\tE_u[ \Phi(u^{-1}k) ]$ where $\Phi(\cdot)$ is the standard normal c.d.f., 
$u$ is the random scale parameter and $\tE_u[\cdot]$ denotes expectation with respect to its distribution. Thus $\pi(m) = 1- \tE_u[\Phi(u^{-1}m)]$.

The above theory of the normal scale mixture structure $g(k)$ and $\pi(k)$  further implies that 
$\emr =  1- \tE_{v,u}[ \int_k\Phi(u^{-1}k)v^{-1}\phi\{v^{-1}(k-m)\} dk]$ with expectation over $v,u$ (in which, implicitly, $u\ci v)$.
Routine normal theory  yields 
$\emr = 1- \tE_{v,u}[\Phi(w^{-1}m)]$ where $w=\sqrt{v^2+u^2}.$     
Hence $\tE_k[\pi(k)] - \pi(m) = \tE_{v,u}[\Phi(u^{-1}m) - \Phi(w^{-1}m)].$   Now, $m\ge 0$ and $w>u$ so that 
$\Phi(u^{-1}m) - \Phi(w^{-1}m) \ge 0$  implying that $\tE_k[\pi(k)] \ge \pi(m),$  as required.  This inequality is strict unless $k=v=0$.  

The analysis above may extend to more general cases when $g(k)$ is not symmetric.  Suppose, for example, that $g(k)$ is a scale mixture of skew-normal distributions~\citep{Azzalini2013}.  This is a rich class of unimodal distributions with ranges of asymmetries; it includes the above symmetric distributions as special cases.  Convolutions of skew-normals with normals are skew-normals, so it is reasonable to ask if the above development generalizes. There may be broader generalization so long as $g(k)$ is unimodal with  $m>0$ and/or $\tPr(k\ge 0)>0.5$. This is an aside and beyond current scope, but suggests further theoretical study.   
Importantly, the above discussion does not extend at all to cases-- including many practical cases-- when the expectation of $k$ (defining the KL divergence) does not exist and when the distribution of $k$ is less regular and even multimodal. 

As another aside note, this also provides interpretation of KL on the probabilistic concordance scale in cases when the bound is assured; in such cases, $\kpf,\kfp \le \log\{(1-\pi_{pf})/\pi_{pf}\}$.

\subsection{Approximation and Bounds\label{app:LinApproxEMR}} 

 The link of EMR to KL is further illuminated in the 1st-order Taylor series approximation of the function $1/\{1+\exp(k)\}$  at $k=0,$ namely
 $1/\{1+\exp(k)\} \approx (2-k)/4$.  This is an exact lower bound on $1/\{1+\exp(k)\}$  for $k\ge 0$ and an exact upper bound for $k\le 0.$
 See this as follows. 
First, $\pi(k) -(2-k)/4$ is positive on $k>0$ if, and only if, $g(k)>0$ where $g(k) = k+2-(2-k)\exp(k).$   
Calculus shows that $g(k)$ is strictly increasing for all $k,$ and, of course, $g(0)=0,$  hence the lower bound arises for $k>0$.
Second,  $\pi(k)  -(2-k)/4$ is negative on $k<0$ if, and only if, $g(k)<0$, so the upper bound is implied on that range.
This approximation is very accurate over $|k|\le 0.5$ where $1/\{1+\exp(k)\} \ge 0.38,$ i.e., in  
cases of relatively close concordance; the absolute error is less than 
$0.68\%$ on $|k|\le 0.5$.  Hence, if $\y\sim \py$ and the implied distribution of $k(\y)$  heavily favors 
 such ranges, then $\emr \approx \{2-\kpf\}/4.$  On this basis, choosing $f(\cdot)$ to maximize $\emr$ is 
 again approximately the (symmetrized) KL divergence minimizing solution.  In the following example highlights  values of
 $\emr\ge 0.4$ as practically relevant, as lower values are strong indications of lack of concordance.

\subsection{A Simple Example\label{sec:calibEMRkluninormalEG} }

A simple example relates the range of $\emr$ to the familiar, interpretable measure of expected sample size (ESS) from Monte Carlo (MC) analysis using importance sampling (IS), in addition to KL. 
Take $\y=y$ to be scalar with $p(y) =\tN(0,1),$  standard normal, and $f(y)=\tN(a,1)$ for some mean 
$a\ge 0.$  IS with proposal $p(y)$ and target $f(y)$ as target leads to MC integration based on the resulting weighted average approximations.  A random sample $y_i\sim p(y)$, $(i=\seq 1n)$, leads to   IS weights $w_i \propto f(y_i)/p(y_i)$ subject to summing to 1.  The resulting MC approximation  to EMR is 
 $\emr\approx \sum_{i=\seq 1n} w_i/(1+nw_i).$ The IS effective sample size as a percentage is
 $\textrm{ESS}=n^{-1}100/\sum_{i=\seq1n} w_i^2$.  Also, in this simple example $\klpf=\klfp=\kpf=a^2/2.$ 
  
\begin{figure}[htbp!]
\centering
\includegraphics[clip,width=3.2in]{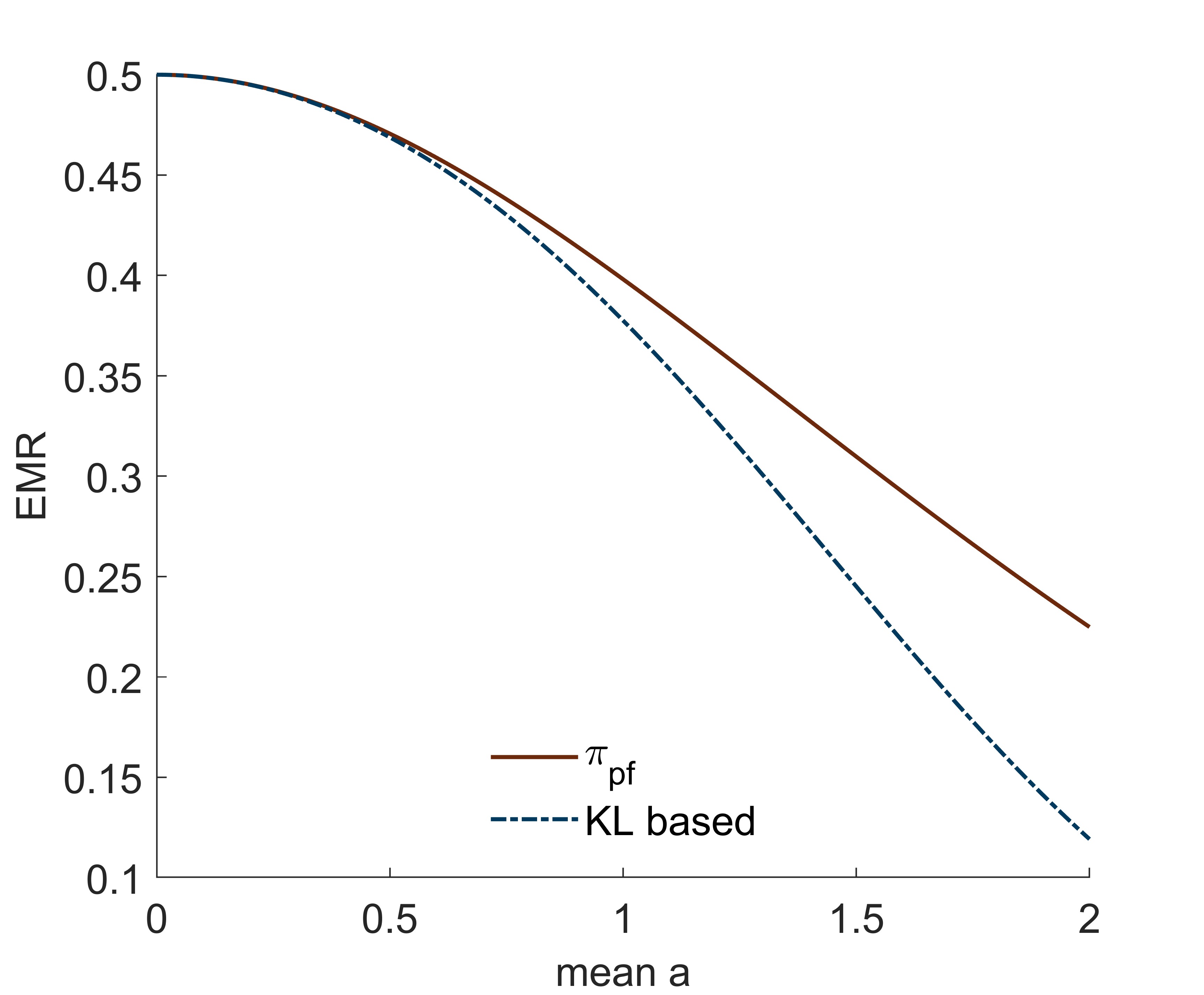}  
\includegraphics[clip,width=3.2in]{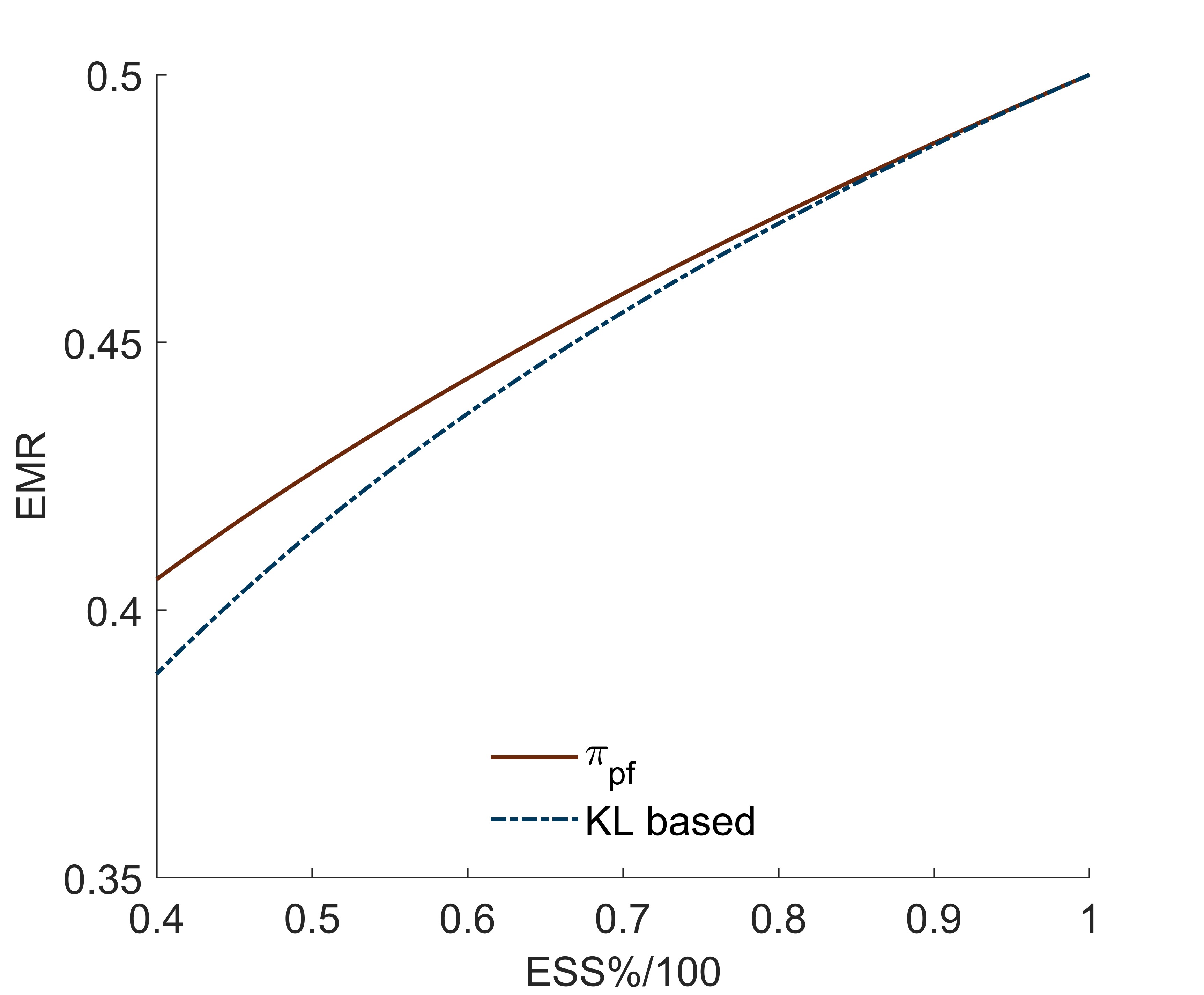}  
\caption{Predictive Concordance Example: EMR, ESS and KL-based lower bound when $p(y)=\tN(0,1)$ and $f(y)=\tN(a,1)$ for a range of values of $a$.
 \label{fig:piessnormalEG}}
\end{figure}

Figure~\ref{fig:piessnormalEG} comes from an example with $n=10^6$ and across a range of values of $a>0$.  
For $a\le 1,$  $\textrm{ESS}\ge 40\%, \emr\ge 0.40$  and is roughly linear in ESS up to its maximum of 0.5. For practical purposes and
extrapolating from this interpretable example-- also supported by other empirical examples--  $\emr\ge 0.4$ or so is expected unless $f(\cdot)$ and $p(\cdot)$ are quite substantially discordant.
From Section~\ref{sec:EMRandKL}, $\emr\ge 0.40$  links to the accurate approximation $\pi_{pf} \approx 1/\{1+\exp(\kpf)\}$ with $\kpf\le 0.4.$

\FloatBarrier

\section{Prior/Regularization Parameter Specification\label{sec:calibrateepsilon} } 

We have prior $\balpha\sim\tDir(\bone(1+\epsilon))$ and aim to calibrate the choice of small $\epsilon.$   We discuss this by analogy with canonical setting for Dirichlet prior/posterior distributions, i.e., multinomial sampling. With $\balpha$ representing scenario probabilities, the least informative multinomial sample is just one draw from one of the scenarios; the implied Dirichlet posterior 
then has parameter updated by $+1$ in one element only. 
With no loss of generality, suppose a single outcome is known to come from the baseline; this single draw posterior is then $\tDir(\bone(1+\epsilon)+\e)$ where $\e=(1,0,\ldots,0)'$.  Now, 
to reflect a minimally informative setting, suppose the posterior is modified to $\tDir(\bone(1+\epsilon)+x\e)$ 
for some very small, positive $x$. This can be regarded as the posterior under an imaginary fractional observation; for example, $x=0.01$ says the information content of the posterior relative to the prior is 1\% of that arising on observing a single multinomial draw.

Under this posterior with specified $x,$ the prior mode $1/(J+1)$ increases to posterior mode $\alpha_0^*=(\epsilon+x)/\{(J+1)\epsilon+x\}$ on $\scn_0$, and decreases to  $\alpha_j^*=\epsilon/\{(J+1)\epsilon+x\}$ on the other scenarios $j>0.$ In this minimal information context it is rationale to limit this latter \lq\lq shrinkage towards zero" and we reflect this by asking that $\alpha_j^*\ge p/(J+1)$ for some fractional reduction $p\in (0,1).$  This implies $\epsilon\ge cx/(J+1)$ where
$c=p/(1-p)$.  Here $c$ is explicitly a lower bound on the reduction from prior to posterior {\em odds} on $\scnj$ for $j>0$ given the minimal information of a single outcome under $\scn_0.$  For example, the choice $c=0.5$ limits this odds reduction to no more than 50\%.   The choices $x=0.01$ and $c=0.5$  imply $\epsilon\ge 0.005/(J+1),$ and this value is recommended as a default.

\section{Optimization of Scenario Mixtures\label{app:MLEconvex}}

For any $\y$ define the $(J+1)-$vector $\p(\y)=[p_0(\y),\ldots, p_J(\y)]'$. It is then easily shown that derivatives of EMR in~\eqn{Eppyfalpha} are
$$ 
  \h(\balpha) \equiv \frac{\delta\emr(\balpha)}{\delta\balpha} = \int_\y \p(\y) h(\y|\balpha) d\y
\quad \textrm{and} \quad
\H(\balpha) \equiv \frac{\delta^2\emr(\balpha)}{\delta\balpha\balpha'}
= -2\int_\y \p(\y)\p(\y)' H(\y|\balpha) d\y
$$
where $h(\y|\balpha) = \py^2/\{\py+\fya\}^2$ and $H(\y|\balpha) = h(\y|\balpha)/\{\py+\fya\}.$  
At any $\balpha$ the Hessian matrix is  $\H(\balpha) = -2 \,\tE[ \p(\y)\p(\y)'  a(\y|\balpha) ]$ where 
$a(\y|\balpha) = \py/\{\py+\fya\}^3$ and the expectation is with respect to $\y\sim p(\cdot).$  
Since $a(\y|\balpha)>0$ for all $\y,\balpha$ the expectation is a positively weighted average of  rank-one 
matrices $\mathbf{p}(\y)\mathbf{p}(\y)'.$ Whether  $\y$ is continuous or discrete (the latter with at least $J+2$ support points) 
$\H(\balpha) $ is full rank and strictly negative definite for all $\balpha.$ 

It follows that maximizing $\emr(\balpha)$ over the simple is a convex optimization problem with a unique maximizing value $\alh$;   standard constrained optimization algorithms then apply.    Further, the modification to add the prior penalty and define the log posterior objective function in~\eqn{Logpostalpha} maintains convexity and ensures a unique posterior mode $\als$ given any specified value of $\epsilon.$  
Then, standard constrained, non-linear optimization methods~(e.g., the default interior-point algorithm in the {\tt fmincon} function in~\citealp{MatlabOPT}) apply and are fast and efficient.

As an aside but of some broader interest, 
the same approach shows that minimizing $\klpf$ or $\klfp$ (when finite) with respect to $\balpha$ are also convex optimizations with unique solutions and similar characteristics. This also applies with any value of $\epsilon$ in the fully Bayesian version that adds the prior penalty based on a very diffuse but proper Dirichlet prior that supports non-zero $\alpha_j$ with probability one.  This links to an existing literature on sparsity and stability of KL-optimal mixtures in the context forecast combination~\citep[e.g.][]{conflitti2015optimal,DIEBOLD2023,crump2024changing,de2024multiplicative}.  Then, relative to EMR, the KL analysis typically leads to more zeros among the optimizing values of $\balpha$ i.e., a sparser mixture more aggressively favoring  just one or a small number of scenarios.  This arises since KL involves expectations of the \bem{unbounded} function $\log\{\py/\fya\}$ and is very dependent on behaviour of the tails of the two p.d.fs.  This also relates to the caveat that, as noted in Section~\ref{sec:EMRandKL}, 
KL divergence may simply be undefined in important practical contexts depending on the relative tail behavior of $\py$ and $\fya.$

In contrast, EMR is more conservative (and numerically more robust) in discounting scenarios that are less concordant with the reference though not extremely so; this arises as EMR is based on expectations of the \bem{bounded} function $1/\{ 1+\py/\fya\}$ in~\eqn{Eppyfalpha}.    That said, the full shrinkage to boundaries of the simplex still arises and requires modest regularization as provided by the penalty induced under the minimally informative prior in the foundational Bayesian setting of Section~\ref{sec:priorposteriorEMR}.

\section{Summary of Computational Flow\label{sec:summaryofflow}}
\begin{enumerate}
	\item Generate a large random sample $\y^i$, $(i=\seq 1n)$, from the reference $p(\y)$, to define an importance sample for Monte Carlo evaluation of the baseline $p_0(\y)$ and the scenarios $p_j(\y).$ 
	\item Evaluate baseline  IS weights $w_0^i \propto p_0(\y^i)/p(\y^i)$, subject to normalization. 
	\item Evaluate the scenario p.d.f.s $p_j(\y)$ for $j>0.$ 
    \begin{enumerate}
        \item If the scenario p.d.f.s are completely specified and can be evaluated, this is as in Step 2 above now applied to scenario p.d.f.s $p_j(\y)$ instead of the baseline $p_0(\y).$ For each $\scnj$ this delivers normalized IS weights $ w_j^i $ on the reference sample values. 
        \item If the scenarios are only partially specified, proceed as follows. 
        \begin{enumerate}
            \item Compute the scenario distributions using a random sample $\x^i$, $(i=\seq 1n),$  
             drawn from the baseline $p_0(\x)$. Use this for MC evaluations of the integrals required to deliver tilting parameters for each $\scnj$ to minimally distort the baseline to match  specified  scenario medians, percentiles, etc.   
           \item Compute the implied  tilting weights $u^i_j$ on each of the reference sampled values $\y^i.$ 
            \item Compute the implied ET-IS weights for each $p_j(\y^i)$ at the reference random sample values $\y^i,$ namely the normalized weights 
            $ w_j^i \propto w_0^iu_j^i$, $(i=\seq 1n),$ for each $\scnj.$ 
        \end{enumerate}
    \end{enumerate}
	\item Compute synthesis weights by optimizing \eqn{Logpostalpha} over $\balpha$; this is modified with constraint $\alpha_0\ge \max_{j=\seq 1J} \alpha_j$ if the baseline is required to be the modal scenario as used in our examples.  At each value of $\balpha$ in iterations of a numerical optimization routine, direct MC integration evaluates EMR in \eqn{Eppyfalpha}. This is just the average over $i=\seq 1n$ of 
    sampled EMR values $ w_f^i(\balpha)/ \{w_f^i(\balpha) + w_p^i\}$ 
    with  implied scenario synthesis IS weights $w_f^i(\balpha) = \sum_{j=\seq 0J} \alpha_j w_j^i$
     and uniform reference weights $w_p^i=1/n$  for all $i=\seq 1n.$

\end{enumerate}

\end{appendices}

\end{document}